\documentclass[journal]{IEEEtran}
\usepackage{comment}
\usepackage{amsmath}
\usepackage{multirow}
\usepackage{color}
\usepackage{caption}
\usepackage{subcaption}
\usepackage{color}
\usepackage{times}
\usepackage{hyperref}
\hypersetup{
	colorlinks=true,
	linkcolor=blue,
	filecolor=blue,
	urlcolor=blue,
	citecolor=cyan,
}
\usepackage{graphicx}
\usepackage{amsmath}
\usepackage{amssymb}

\ifCLASSINFOpdf
\else
\fi
\begin{document}
%
\title{Lightness Modulated Deep Inverse Tone Mapping}
%
%
%

\author{Kanglin Liu, Gaofeng Cao, Jiang Duan, Guoping Qiu
\thanks{Kanglin Liu is with Pengcheng Laboratory, P.R.China, e-mail: max.liu.426@gmail.com}
\thanks{Guoping Qiu is with University of Nottingham and Shenzhen University, e-mail: guoping.qiu@nottingham.ac.uk}
\thanks{Manuscript received *, *; revised *,*.}}

%
%

\markboth{}%
{Shell \MakeLowercase{\textit{et al.}}: Bare Demo of IEEEtran.cls for IEEE Journals}
%



\maketitle

\begin{abstract}
Single-image HDR reconstruction or inverse tone mapping (iTM) is a challenging task. In particular, recovering information in over-exposed regions is extremely difficult because details in such regions are almost completely lost. In this paper, we present a deep learning based iTM method that takes advantage of the feature extraction and mapping power of deep convolutional neural networks (CNNs) and uses a lightness prior to modulate the CNN to better exploit observations in the surrounding areas of the over-exposed regions to enhance the quality of HDR image reconstruction. Specifically, we introduce a Hierarchical Synthesis Network (HiSN) for inferring a HDR image from a LDR input and a Lightness Adpative Modulation Network (LAMN) to incorporate the the lightness prior knowledge in the inferring process. The HiSN hierarchically synthesizes the high-brightness component and the low-brightness component of the HDR image whilst the LAMN uses a lightness adaptive mask that separates detail-less saturated bright pixels from well-exposed lower light pixels to enable HiSN to better infer the missing information, particularly in the difficult over-exposed detail-less areas. We present experimental results to demonstrate the effectiveness of the new technique based on quantitative measures and visual comparisons. In addition, we present ablation studies of HiSN and visualization of the activation maps inside LAMN to help gain a deeper understanding of the internal working of the new iTM algorithm and explain why it can achieve much improved performance over  state-of-the-art algorithms.
\end{abstract}


%
\IEEEpeerreviewmaketitle

\section{Introduction}
In contrast to low dynamic range (LDR) imaging, high dynamic range (HDR) imaging is able to capture, manipulate and display real-world scenes \cite{banterle2017advanced,banterle2009high,Reinhard2010hdrimaging}. 
HDR imaging has promising applications in photography, physically-based rendering, gaming, films, medical imaging, and improving the viewing experience \cite{Eilertsen2017review,Kalantari2017deep}. 
Unfortunately, the status quo is that the majority of both current and legacy content is predominantly LDR,  hence inducing the growing demand for converting LDR to HDR \cite{Debevec1997recovering,Mertens2009exposure,Pan2020multi}.

The most common approach to generating an HDR image is to merge multiple LDR images captured with different exposures. Such a technique performs well on static scenes but often suffers from ghosting artifacts on dynamic scenes or hand-held cameras. Furthermore, capturing multiple images of the same scene may not always be feasible \cite{Fotiadou2019,Zhang2018,Xu2019}.

Single-image HDR reconstruction, referred to as inverse tone mapping (iTM), aims to recover an HDR image from a single LDR input.
iTM can be achieved by model-based or learning-based methods.
Model-based methods try to recover an HDR image by introducing various tone mapping operators (TMOs), which rely on different prior knowledge \cite{Huo2013,Huo2014physiological,Kinoshita2017fast,Kovaleski2014,Masia2009}. Though model-based methods are usually algorithmically interpretable, they are intensively parameter dependent, making them user-unfriendly for non-experts and unsuitable for all types of contents \cite{Pan2020multi,Zhang2018}.
\begin{figure*}[htp]
	\centering
	\includegraphics[height=0.15\linewidth]{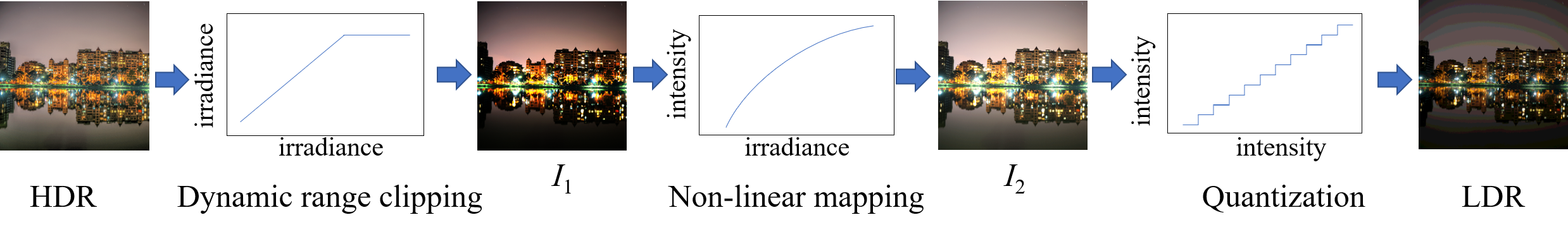}	
	\caption{LDR imaging pipeline. Firstly, an HDR image would be clipped to the range of [0, 1], resulting in the missing contents in the over-exposed regions. Secondly, non-linear mapping with a camera response function would transfer a linear irradiance to a non-linear brightness. Thirdly, the recorded pixels are quantized to 8-bites, leading to the visual artifacts mainly in the under-exposed regions.}
	\label{fig:fig1}
\end{figure*}

With the development of deep convolutional neural networks (CNNs), learning-based methods tackle the iTM problem by learning a mapping function from LDR images to their HDR estimations, which is generally supervised by given LDR/HDR pairs. However, the variation of HDR pixels (32-bits) is significantly higher than that of LDR pixels (8-bits), resulting in difficult LDR-to-HDR mapping \cite{Zhao2019deep}.
To address this challenge, literature \cite{Eilertsen2017hdr} only recovers the over-exposed regions, simplifying the mapping function but  decreasing the visual quality as well.  Endo \textit{et al.} generate multiple up-/down-exposed LDR images and fuses them to synthesize an HDR image \cite{Endo2017deep}. Some attempt to improve the iTM performance via carefully designed networks, \textit{e.g.}, Marnerides \textit{et al.} have introduced the ExpandNet \cite{Marnerides2018expandnet}, and Lee \textit{et al.} have applied Generative Adversarial Networks (GANs) \cite{Lee2018hdri}.
Moreover, Liu \textit{et al.} incorporate the prior knowledge of the LDR imaging pipeline, and decompose the iTM problem into three sub-tasks \cite{Liu2020single}. To achieve this, three cascaded networks are utilized to remove artifacts caused by quantization, camera response functions (CRFs) process and sensor saturation, respectively. 
To infer the missing information, nearly all learning-based methods convert LDR to HDR via stacking convolutional and nonlinearity layers. 
However, conventional convolution operations are not suitable for restoring details in regions of different lightness as they are spatially equivalent. Specifically, missing details caused by sensor saturation only occur in the over-exposed regions, while a standard convolutional operation applies identical filters across all pixels, in a sliding window manner \cite{Kim2020jsigan}. 
Clearly, in order to improve the quality of HDR image reconstruction, image regions should be treated differently. In particular, it is the very bright areas of the scene that need special attention because it is where image details are often missing due to sensor saturation.

To this end, a Hierarchical Synthesis Network (HiSN), that incorporates prior knowledge of LDR image formation pipeline, is proposed to hierarchically synthesize the dim part  $H_1 \in [0, 1]$ and the bright part $H_2 \in (1, + \infty)$ of the image. 
To be specific, $H_1$ will be responsible for correcting camera non-linear response to recover a linear irradiance, and for compensating the missing information caused by quantization.
Meanwhile, $H_2$ is responsible for estimating the saturated pixels in the over-exposed regions. The final HDR estimation is generated by combining $H_1$ and $H_2$.
In addition, we introduce the Lightness Adaptive Modulation Network (LAMN), which uses a lightness adaptive mask to mark out over-exposed high lightness areas where pixels are saturated and there is no or little details, and areas of  lower lightness where there is detail.
Then, LAMN would use scaling and bias terms to adaptively and discriminately modulate the features in HiSN. Thus, HiSN can focus on relevant features when inferring the saturated pixels.
To summarize, our contributions are as follows:

\noindent(1) We have developed a new state of the art inverse tone mapping method consisting of a Hierarchical Synthesis Network (HiSN) and a Lightness Adaptive Modulation Network (LAMN). HiSN constructs the high-brightness component and the low-brightness component of the image in a hierarchical manner while LAMN uses lightness prior knowledge to modulate the features maps of HiSN to enable it to process different lightness regions discriminately to improve its performance in recovering the missing HDR information, particularly in the difficult over-exposed detail-less regions. 
%

\noindent(2) We provide extensive quantitative and qualitative experimental results to demonstrate that our new iTM technique outperforms state of the art algorithms.


\noindent(3) We provide insights into the working of HiSN and LAMN to help explain how they works and why they can achieve better iTM performances than similar work in the literature.


\section{Related Work}
In contrast to the real scene irradiance, which usually has a high dynamic range, the camera sensor can only capture and record a limited extent \cite{Lee2018hdri,Zhang2017learning}. Given the scene irradiance $E$ and sensor exposure time $t$, an HDR image $H^t$ is stored as: $H^t = E \times t$. As shown in Fig. \ref{fig:fig1}, the LDR imaging pipeline converts one HDR image to the corresponding LDR one, which can be modeled by the following major steps \cite{Kalantari2017deep,Debevec1997recovering,Liu2020single}:

\noindent(1) Dynamic range clipping. Due to sensor saturation, the pixel values of an HDR image would be clipped to a limited range, which can be formulated by: $I_1=\mathcal{C}(H^t)=min(H^t, 1)$, where $\mathcal{C}$ represents the dynamic range clipping process, and $H^t$ is the ground-truth HDR image. Owing to this, the details in the over-exposed regions would disappear.

\noindent(2) Non-linear mapping with a camera response function (CRF). In order to match the human visual system, a non-linear mapping function $\mathcal{F}$ is typically applied to convert a linear scene irradiance to a non-linear brightness: $I_2= \mathcal{F}(I_1)$. CRFs are determined by the camera models, and the Database of Response Functions (DoRF) has collected 201 CRF curves for common brands of films, charge-coupled devices (CCDs), and digital cameras \cite{Grossberg2003}.

\noindent(3) Quantization. The recorded pixel values are quantized to 8-bits by $\mathcal{Q}(I_2) =[I_2 \times 255+0.5]/255$, and quantization would lead to visual artifacts in under-exposed regions.
Overall, the LDR imaging pipeline can be formulated as:
\begin{equation}\label{key1}
	L=\mathcal{Q}(\mathcal{F}(\mathcal{C}(E \cdot t)))
\end{equation}

To successfully recover the HDR image from an LDR image, single-image HDR reconstruction or inverse tone mapping (iTM)  algorithms need to infer the missing contents caused by sensor saturation and quantization, and calibrate the LDR image to obtain a linear irradiance. Conventionally, iTM algorithm is implemented via model-based or learning-based methods.

\begin{figure*}[htp]
	\centering
	\includegraphics[height=0.48\linewidth]{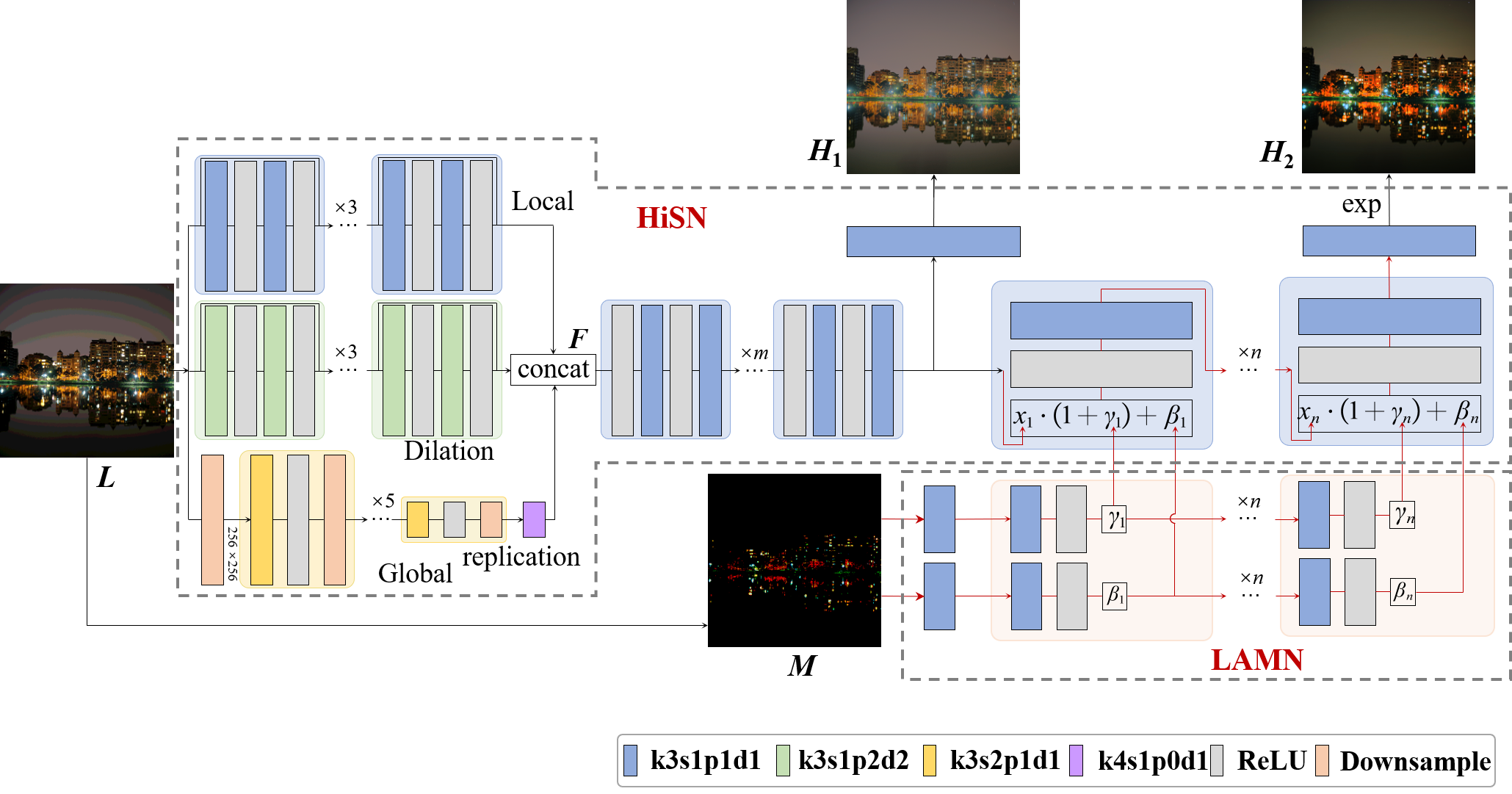}	
	\caption{An overview of HiSN and LAMN. Givn LDR inputs $L$, HiSN is responsible for synthesizing the dim part $H_1$ first, then estimates the bright part $H_2$ with the modulation of LAMN, which utilizes the mask $M$ to adaptively and discriminately modulate the activation maps $x_i$ in HiSN. k3s1p1d1 denotes a convolutional layer with kernel size 3, stride 1, padding 1 and dilation 1. The same applies to k3s1p2d2, k3s2p1d1 and k4s1p0d1. ReLU is the non-linear layer, and Downsample is the average pooling layer. $m$ and $n$ is empirically taken as 5 and 6, respectively.}
	\label{fig:fig2}
\end{figure*}

\noindent{\textbf{Model-based method}} relies on different prior knowledge to obtain the inverse tone mapping operators (iTMOs) for converting LDR to HDR \cite{banterle2009high,Mertens2009exposure,Huo2013,Huo2014physiological,Kinoshita2017fast,Kovaleski2014,Masia2009}. Specifically, Huo \textit{et al.}  have introduced the iTMO according to the human visual system and its retina response \cite{Huo2014physiological},  and Aky{\"u}z \textit{et al.} established the mapping function using the gamma functions, which is capable of expanding the dynamic range \cite{Akyuz2007hdr}.
While Kovaleski \textit{et al.} utilized an edge-preserving expansion map from bilateral grids to enhance the details in bright regions\cite{Kovaleski2014high}.
It is clearly seen that model-based methods are algorithmically interpretable \cite{Kinoshita2017fast}. However, those iTMOs involve in numerous parameters, making them user-unfriendly for non-experts and unsuitable for all types of contents. 
Most importantly, existing operators focus on boosting the dynamic range to make it look plausible on an HDR display, or to produce rough estimates needed for image-based lighting.
The resulting HDR images tend to have poor visual quality especially in over exposed regions \cite{Eilertsen2017hdr,Marnerides2018expandnet}.

\noindent{\textbf{Learning-based method}} has attracted intensive attentions recently. In contrast to model-based method, learning-based approaches utilize convolutional neural networks (CNNs) to learn the mapping function from LDR images to their HDR estimations. 
Considering the difficulty of converting 8-bits LDR to 32-bits HDR, different approaches have introduced strategies to tackle the iTM problem. 
Specifically, Eilertsen \textit{et al.} have introduced a U-Net architecture, termed HDRCNN \cite{Eilertsen2017hdr}, to predict the saturated pixels, ignoring the under-exposed regions, then synthesized the final HDR by combining the predicted saturated pixels and linearized LDR images, which are calculated by a handcrafted inverse CRF, \textit{i.e.}, $\mathcal{F}^{-1}(x)=x^{2}$. 
By contrast, Endo \textit{et al.} predicted multiple exposures  from a single exposure which were then used to generate an HDR image using standard merging algorithms \cite{Endo2017deep}. 
More explicitly, Liu \textit{et al.} incorporated the prior knowledge of LDR imaging, and modeled the LDR imaging pipeline, then applied three cascaded networks to model the inverse functions of the quantization, non-linear mapping, and dynamic range clipping, respectively \cite{Liu2020single}. 
In order to improve the iTM performance, Marnerides \textit{et al.} have introduced the ExpandNet \cite{Marnerides2018expandnet}, and Lee \textit{et al.} applied the Generative Adversarial Networks \cite{Lee2018hdri}.
To summarize, the existing methods learn the mapping function under the supervision of LDR/HDR pairs, and the modifications or improvements mainly include carefully-designed CNN architectures and applying the prior knowledge of LDR imaging pipeline. 


\section{Proposed Method}
As indicated by the LDR imaging pipeline, the missing details caused by dynamic range clipping occurs in the over-exposed (high-brightness) parts, while the under-exposed (low-brightness) regions are most likely to suffer from artifacts induced by quantization.
However, almost all CNN based iTM solutions in the literature are based on stacking conventional convolutional-nonlinear layers \cite{Kim2020jsigan} and applied the same operations to all pixels via a sliding window, this makes them potentially unable to infer both missing details caused by dynamic range clipping and by quantization. A better solution would be to treat the two types of missing details, those caused by dynamic range clipping and those caused by quantization, separately by making use of the lightness prior. 

To this end, we introduce a novel iTM solution that combines  a mapping network called Hierarchical Synthesis Network (HiSN) and a lightness prior network called the Lightness Adaptive Modulation Network (LAMN) for recovering a HDR image from a single LDR input.
The following sections will describe the motivation, rationale and design of HiSN and LAMN.

\subsection{Hierarchical Synthesis Network}
Fig. \ref{fig:fig2} provides an overview of the proposed architecture. 
Specifically,  the Hierarchical Synthesis Network (HiSN) is responsible for synthesizing the HDR output from the given LDR input. 
Motivated by \cite{Marnerides2018expandnet}, HiSN utilizes local, dilation and global branches as the feature extraction part. Each one of the three branches is responsible for a particular aspect, with the local branch handling local detail, the dilation branch for medium level detail, and a global branch capturing higher level image-wide features.
A convolutional neural network applies location invariant filtering operations across all pixels in the image, regardless whether it is an over-exposed saturated region without any details or a well-exposed region with sufficient local details. Clearly, the task of recovering HDR information in the very bright areas where they are most likely over-exposed and contain very little details is more difficult than dealing with less bright regions of the image where it is more likely the pixels are better exposed and contain good details. Therefore, simply applying a CNN and treating the very bright saturated regions and less bright non-saturated regions the same way does not makes sense. A better strategy is to treat the very bright regions and the rest of the regions differently because these regions will have to use different kinds of features to recover the lost HDR information. For example, in the very bright regions where the pixels are saturated, there is no information from the regions themselves because all pixels are of the same value, what is needed is to focus on information from their surrounding areas to infer the information in these detail-less areas. Based on these reasonings, HiSN hierarchically generates a bright component $H_2$ and a dim component $H_1$ separately before fusing them together to recover the full HDR image.
To achieve this, the concatenated features $F$ are fused using cascaded convolutional blocks to obtain the dim part $H_1$ first, and then synthesize the bright one $H_2$ via incorporating features from LAMN. 

As indicated by the LDR imaging pipeline in Fig. \ref{fig:fig1} and  (\ref{key1}), the CRF process $\mathcal{F}:[0, 1] \rightarrow [0, 1]$ as well as the quantization process $\mathcal{Q}:[0, 1] \rightarrow [0, 1]$ would not clip or expand the dynamic range. In contrast, the dynamic range clipping process $\mathcal{C}: [0, +\infty ) \rightarrow [0, 1]$ would clip the dynamic range.
Considering that $I_1$ and $I_2$ have identical range to $L$, \textit{i.e.}, $I_1$, $I_2$ and $L$ are all in the range of [0, 1], HiSN first conducts radiometric calibration to obtain a linear irradiance and infers the missing contents caused by quantization simultaneously, thus obtaining the dim part $H_1 \in [0, 1]$. Hierarchically, HiSN then synthesizes the bright part $H_2 \in (1, +\infty)$ for estimating the clipped dynamic range in the over-exposed regions. The final HDR estimation $H$ is obtained by combining $H_1$ and $H_2$:
\begin{equation}\label{key2}
	H=H_1+H_2
\end{equation}

Hierarchically synthesizing the dim and bright part, $H_1$ and $H_2$ is consistence with the inverse process of the LDR imaging pipeline.
Ablation study in Section \ref{hisn} would show the effectiveness of HiSN and its contribution to iTM performance.

\begin{figure}[htp]
	\centering
	\begin{subfigure}[b]{0.23\textwidth}
		\centering
		\includegraphics[height=0.90\linewidth]{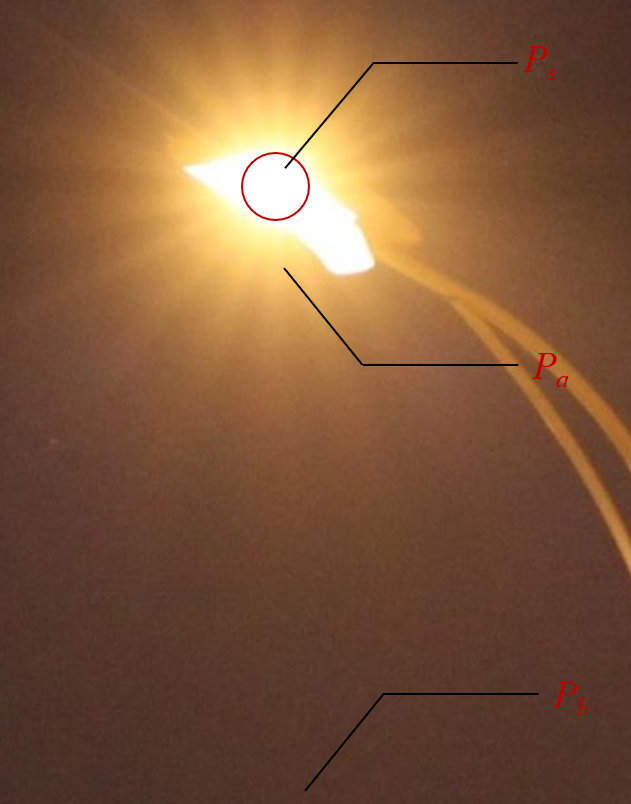}
		\caption{}
	\end{subfigure}	
	\begin{subfigure}[b]{0.23\textwidth}
		\centering
		\includegraphics[height=0.90\linewidth]{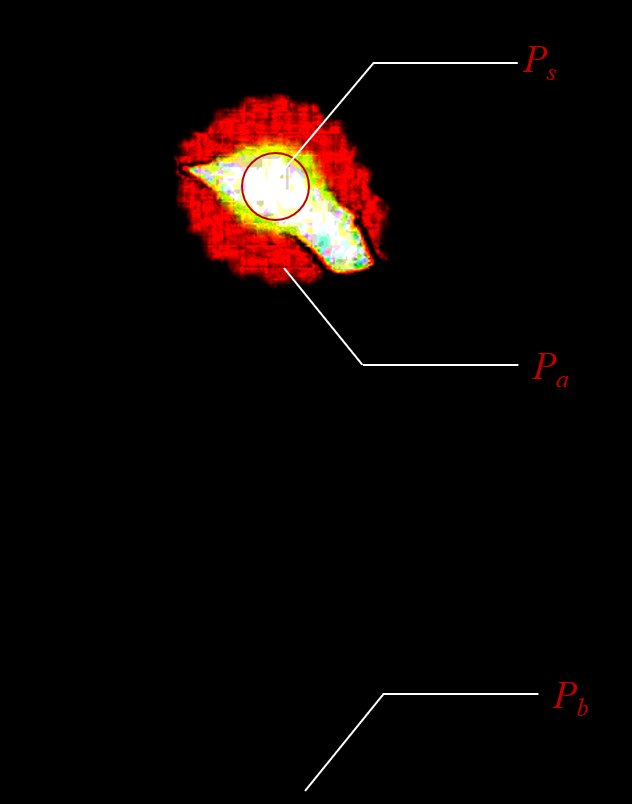}
		\caption{}
	\end{subfigure}		
	\caption{An example to show the motivation of LAMN. (a) shows a LDR image with limited dynamic range. The lamp region marked by the red circle suffers from sensor saturation, resulting in the saturated pixels $P_s$. $P_a$ and $P_b$ are the pixels around and far away from $P_s$, respectively. The lamp has determined the irradiance intensity, thus deciding the recorded pixel values of $P_s$ and $P_a$. However, $P_b$ is hardly affected by the lamp, as $P_b$ is far away from the lamp. (b) shows the mask of (a), which is obtained via (\ref{key3}). It is clearly seen that the mask distinguishes $P_s$ and $P_a$ from $P_b$.}
	\label{fig:fig3}
\end{figure}
\subsection{Lightness Adaptive Modulation Network}
Due to sensor saturation, pixel values in the over-exposed regions would be clipped to the range of $[0, 1]$. Thus, inferring the missing details in the over-exposed regions becomes the most critical step of the iTM task.
Let $P_s$ denote the saturated pixels in the LDR images, $P_a$ denote pixels around $P_s$, and $P_b$ denote pixels far away from $P_s$.
For conventional learning-based methods, convolutional kernels are applied across all the pixels in a sliding window manner, which is at best sub-optimal. Because $P_s$ and $P_a$ will have more relevance to inferring the scene irradiance in the over-exposed regions than $P_b$.
To elaborate this, a simple example is presented in Fig. \ref{fig:fig3}, where image contrast disappears in the lamp region marked by the red circle. 
The lamp in Fig. \ref{fig:fig3} is an irradiance source of the scene, which in turn influences the pixel values of $P_s$ and $P_a$, while $P_b$ will be much less correlated with $P_s$ because it is far away from the lamp.
Thus, to infer the scene radiance in the over-exposed regions, we are supposed to pay more attention to $P_s$ and $P_a$ than $P_b$.
However, existing methods treat these pixels equally, and ignore these important differences.
To address this, the Lightness Adaptive Modulation Network (LAMN) is proposed to work alongside HiSN for restoring the missing content in the over-exposed regions.

Let $M$ denote a mask:
\begin{equation}\label{key3}
	M=\frac{max(0, L-\tau )}{1-\mathsf{\tau}}
\end{equation}
where $\tau$ is a threshold value, determining how many pixels around $P_s$ should be considered. Empirically, $\tau$ is taken as 0.95 in this paper. Fig. \ref{fig:fig3}(b) show an example of $M$, which clearly distinguishes $P_s$ and $P_a$ from $P_b$.

\begin{table*}[htp]
	\centering
	\caption{Quantitative evaluation of iTM performance. PU represent the perceptual uniformity encoding, PSNR, SSIM, MS-SSIM are  Peak Signal to Noise Ratio, Structural Similarity and Multi-Scale Structural Similarity, respectively. The bold indicates the best result.}
	\resizebox{\textwidth}{20mm}{
		\begin{tabular}{l|c c c c||c c c c|| c c c c}
			\hline\hline
			&\multicolumn{4}{c||}{HDR-SYNTH}&\multicolumn{4}{c||}{HDR-REAL}&\multicolumn{4}{c}{RAISE}	\\
			& {\scriptsize HDR-VDP} & {\scriptsize PU-PSNR} & {\scriptsize PU-SSIM} & {\scriptsize PU-MS-SSIM} & {\scriptsize HDR-VDP} & {\scriptsize PU-PSNR} & {\scriptsize PU-SSIM} & {\scriptsize PU-MS-SSIM} & {\scriptsize HDR-VDP} & {\scriptsize PU-PSNR} & {\scriptsize PU-SSIM} & {\scriptsize PU-MS-SSIM}\\
			\hline
			AEO \cite{Akyuz2007hdr}&50.88&19.75&0.27&0.21&57.90&30.24&0.56&0.49&57.38&28.78&0.33&0.20\\
			HPEO \cite{Huo2014physiological}&50.91&20.73&0.41&0.32&57.81&31.37&0.63&0.55&59.34&32.18&0.63&0.54\\
			KOEO \cite{Kovaleski2014high}&52.03&23.04&0.43&0.37&59.21&33.60&0.68&0.61&58.51&33.12&0.62&0.55\\
			MEO \cite{Masia2017dynamic}&50.43&20.43&0.32&0.24&54.93&29.19&0.54&0.46&57.57&30.48&0.46&0.34\\
			HDRCNN \cite{Eilertsen2017hdr}&53.97&23.28&0.36&0.29&59.70&33.26&0.65&0.59&59.45&33.17&0.61&0.53\\
			DrTMO \cite{Endo2017deep}&54.62&21.19&0.24&0.19&58.30&27.65&0.51&0.48&60.05&32.45&0.48&0.37\\
			ExpandNet \cite{Marnerides2018expandnet}&51.47&22.21&0.35&0.29&58.58&31.62&0.57&0.53&57.51&32.98&0.60&0.51\\
			SingleHDR \cite{Liu2020single}&57.39&28.04&0.51&0.46&60.81&\textbf{38.29}&\textbf{0.76}&\textbf{0.71}&59.88&30.48&0.60&0.55\\
			Ours&\textbf{57.46}&\textbf{29.3}0&\textbf{0.67}&\textbf{0.58}&\textbf{61.33}&37.95&0.73&0.70&\textbf{60.79}&\textbf{34.43}&\textbf{0.71}&\textbf{0.68}\\
			\hline\hline
	\end{tabular}}
	\label{t1}
\end{table*}

Let $x_i \in R^{C_i \times H_i \times W_i}$ denote the $i$-th activation map in HiSN as shown in Fig. \ref{fig:fig2}, where $C_i$ enumerates the channels, $H_i$ and $W_i$ represent the height and width of the activation map, respectively. LAMN would utilize $M$ to adaptively and discriminately modulate the activation maps $x_i$:
\begin{equation}\label{key4}
	x_{i}^{'}=x_i\cdot(1+\gamma_i)+\beta_i
\end{equation}
where $i=1,\cdots, n$, $x_{i}^{'}$ is the modulated activation map, $\gamma_i$ and $\beta_i$ are determined by:
\begin{equation}\label{key5}
	\gamma_i = \left\{\begin{matrix}
		ReLU(Conv(\gamma_{i-1})) &i>1\\ 
		ReLU(Conv(M)) &i=1
	\end{matrix}\right.	
\end{equation}
\begin{equation}\label{key6}
	\beta_i = \left\{\begin{matrix}
		ReLU(Conv(\beta_{i-1})) &i>1\\ 
		ReLU(Conv(M)) &i=1
	\end{matrix}\right.	
\end{equation}
where $ReLU$ and $Conv$ are the non-linear and convolutional operations, respectively.
It is obvious that each entry in $\gamma_i \in R^{C_i \times H_i \times W_i}$ and $\beta_i \in R^{C_i \times H_i \times W_i}$ has non-negative value.
Fig. \ref{fig:fig2} illustrates the LAMN design, and it is clearly seen that $\gamma_i$ and $\beta_i$ are learned modulation parameters, which depend on the input mask $M$. 
Mathematically, $(1+\gamma_i)$ and $\beta_i$ can be regarded as the scaling and bias terms, respectively, to conduct the linear transformation on $x_i$.
Equivalently, in the context of the inverse tone mapping task, $(1+\gamma_i)$ would enhance the contrast of $x_i$, and $\beta_i$ can add an offset for $x_i$, thus providing a technique for modulating $x_i$.
Interpretation of LAMN and how modulation is conducted can be found in Section \ref{mamn}.


\section{Experiments}
We first describe our experimental settings and evaluation metrics, then present quantitative and visual comparisons with state-of-the-art single-image HDR reconstruction algorithms. 
Following this section, we provide evidences to show the validity of HiSN and LAMN, and explain how modulation is conducted inside LAMN.

\subsection{Experiment Setups}
Following the practices in the literature \cite{Eilertsen2017hdr,Marnerides2018expandnet,Liu2020single}, we utilize the 502 HDR images in the HDR-SYNTH datasets \cite{Eilertsen2017hdr} for training. LDR images are synthesized using the LDR imaging pipeline in  (\ref{fig:fig1}).
Specifically,  we uniformly sample 60 exposure times $t$ in the $log_2$ space within [-3, 3] and apply 171 different CRFs from the training set of the DoRF dataset \cite{Grossberg2003}. 
Identical to that in the literature \cite{Liu2020single}, the Poisson-Gaussian noise is also added in (\ref{key1})  when generating training data, which can  be approximated by a heteroscedastic Gaussian with a signal-dependent variance: $\sigma^2(I)=I\cdot\sigma_{s}^{2}+\sigma_{c}^2$, where $I$ is the pixel intensity, and $\sigma_{s}$, $\sigma_{c}$ are uniformly sampled from the range of [0, 0.013] and [0, 0.005], respectively.
As real-world LDR images often contain JPEG compression artifacts, we save the synthesized LDR images in a JPEG format with a quality factor randomly sampled from the range of [85, 100]. This process can be considered as a data augmentation procedure, which has been shown to be effective in improving the reconstruction quality on real images \cite{Eilertsen2017hdr}.

Accordingly, the loss metric $l$ for training can be formulated by:
\begin{equation}\label{key7}
	l = ||H_1 - \mathcal{C}(H^t)|| + \lambda \cdot ||\mathrm{T}(H_2)-\mathrm{T}(H^t - \mathcal{C}(H^t))||
\end{equation}
where $\lambda$ is the hyperparameter, and $T(\cdot)$ maps the pixel values to their log domain:
\begin{equation}\label{key8}
	\mathrm{T}(H) = \frac{log(1+\mu H)}{log(1+\mu)}
\end{equation}
where $\mu$ is a hyperparameter and usually taken as 5000.
$||H_1 - \mathcal{C}(H^t)||$ would encourage $H_1$ to be identical to $\mathcal{C}(H^t)$, such that $H_1$ is able to conduct radiometric calibration and infer the missing contents due to quantization, and $||\mathrm{T}(H_2)-\mathrm{T}(H^t - \mathcal{C}(H^t))||$ requires $H_2$ to infer the saturated pixels caused by the dynamic range clipping process.
Empirically, measuring the Euclidean distance between $H_2$ and $H^t - \mathcal{C}(H^t)$ in the log domain has seen good performance. Besides, in the literature \cite{Dehaene2003,Portugal2011}, it has been found that measuring the Euclidean distance in the log domain matches the retina response of human visual system (HVS) \cite{Dehaene2003,Portugal2011}, and further contributes to stabilizing the training \cite{Pan2020multi,Marnerides2018expandnet}.
The ADAM optimizer \cite{Da2014method} is used to minimize the loss function of  (\ref{key7}) with initial learning rate of $1 \times 10^{-4}$, which decays exponentially every 5000 iterations. 
Our network is trained with a batch size of 16, and the training is terminated 
when iteration reaches 20,000.

We adopt the  Peak Signal to Noise Ratio (PSNR), Structural Similarity (SSIM), Multi-Scale Structural Similarity (MS-SSIM) \cite{Akyuz2007hdr,Wang2004image,Aydin2008pu} and HDR-VDP-2.2 \cite{Mantiuk2011hdr} to evaluate the accuracy of HDR reconstruction.
For the first three metrics, a perceptual uniformity (PU) encoding \cite{Aydin2008pu} is applied to the prediction and reference images to make them suitable for HDR comparisons. HDR-VDP-2.2 has already included the PU-encoding in its implementation.
Evaluations are conducted on the newly introduced  HDR-REAL dataset \cite{Liu2020single} as well as widely-used ones like the HDR-SYNTH \cite{Liu2020single} and RAISE \cite{Dang2015raise} datasets. 

%

\begin{figure*}[htp]
	\vspace*{-0.4cm}
	\centering
	\begin{subfigure}[b]{0.16\textwidth}
		\centering
		\includegraphics[height=0.875\linewidth]{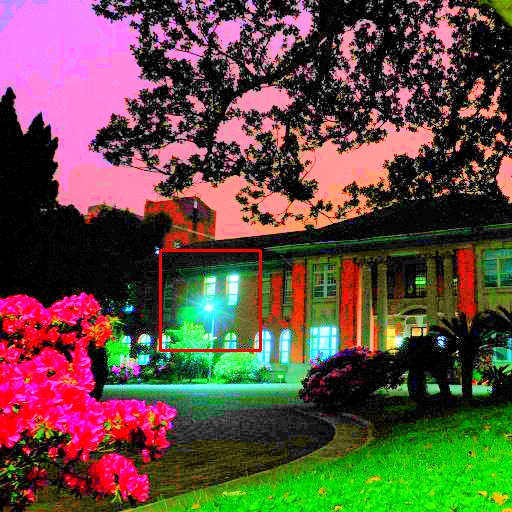}\\
		\includegraphics[height=0.875\linewidth]{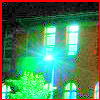}
		\caption{AEO \cite{Akyuz2007hdr}}
	\end{subfigure}
	\begin{subfigure}[b]{0.16\textwidth}
		\centering
		\includegraphics[height=0.875\linewidth]{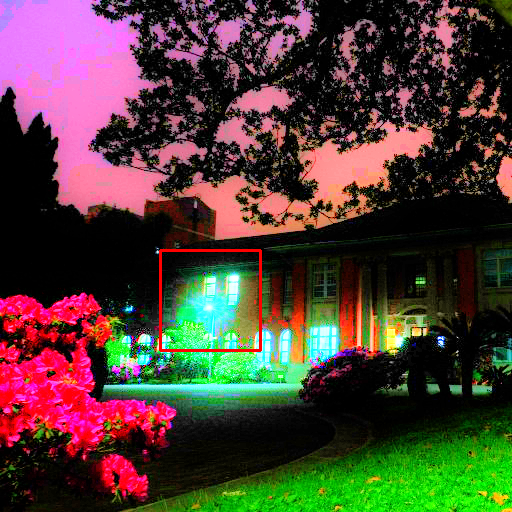}	\\
		\includegraphics[height=0.875\linewidth]{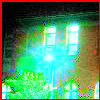}
		\caption{HPEO \cite{Huo2014physiological}}
	\end{subfigure}
	\begin{subfigure}[b]{0.16\textwidth}
		\centering
		\includegraphics[height=0.875\linewidth]{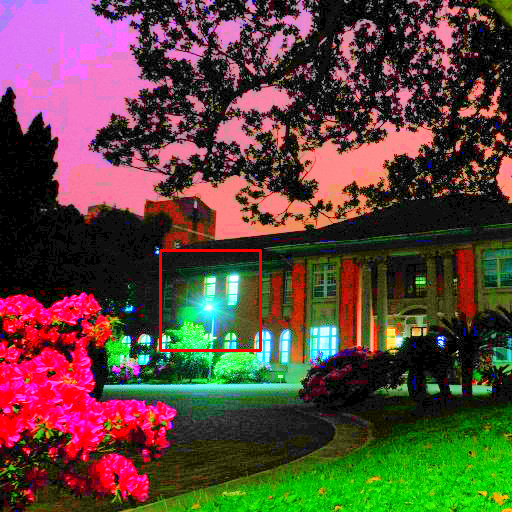}	\\
		\includegraphics[height=0.875\linewidth]{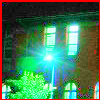}	
		\caption{KOEO \cite{Kovaleski2014high}}
	\end{subfigure}
	\begin{subfigure}[b]{0.16\textwidth}
		\centering
		\includegraphics[height=0.875\linewidth]{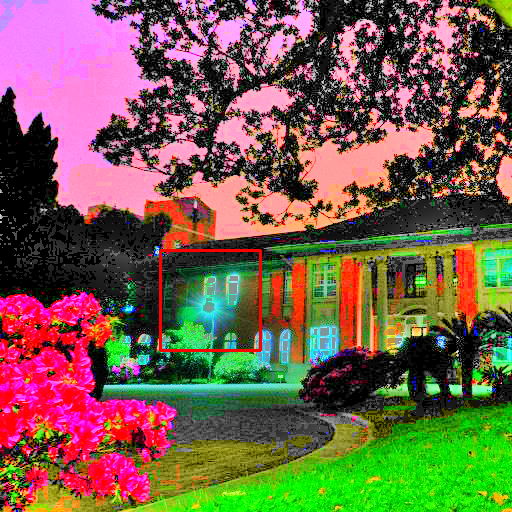}	\\
		\includegraphics[height=0.875\linewidth]{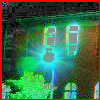}	
		\caption{MEO \cite{Masia2017dynamic}}
	\end{subfigure}
	\begin{subfigure}[b]{0.16\textwidth}
		\centering
		\includegraphics[height=0.875\linewidth]{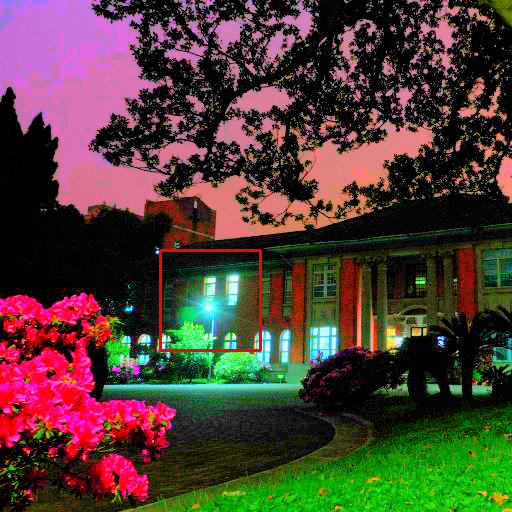}	\\
		\includegraphics[height=0.875\linewidth]{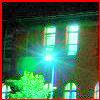}	
		\caption{HDRCNN \cite{Eilertsen2017hdr}}
	\end{subfigure}
	
	\begin{subfigure}[b]{0.16\textwidth}
		\centering
		\includegraphics[height=0.875\linewidth]{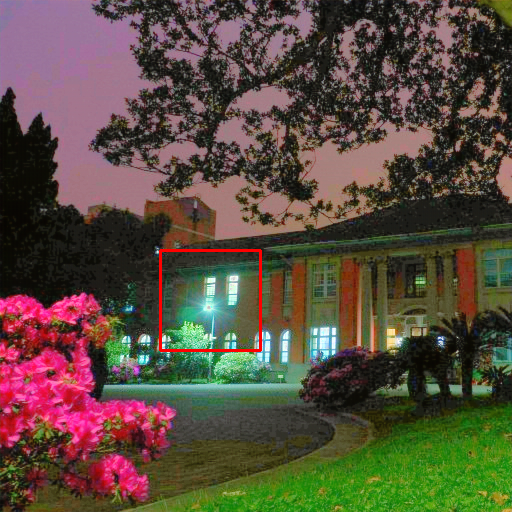}	\\
		\includegraphics[height=0.875\linewidth]{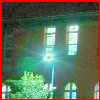}	
		\caption{DrTMO \cite{Endo2017deep}}
	\end{subfigure}
	\begin{subfigure}[b]{0.16\textwidth}
		\centering
		\includegraphics[height=0.875\linewidth]{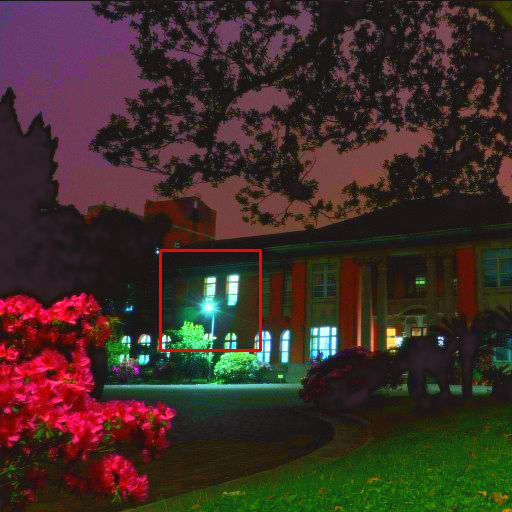}	\\
		\includegraphics[height=0.875\linewidth]{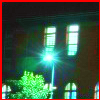}	
		\caption{ExpandNet \cite{Marnerides2018expandnet}}
	\end{subfigure}
	\begin{subfigure}[b]{0.16\textwidth}
		\centering
		\includegraphics[height=0.875\linewidth]{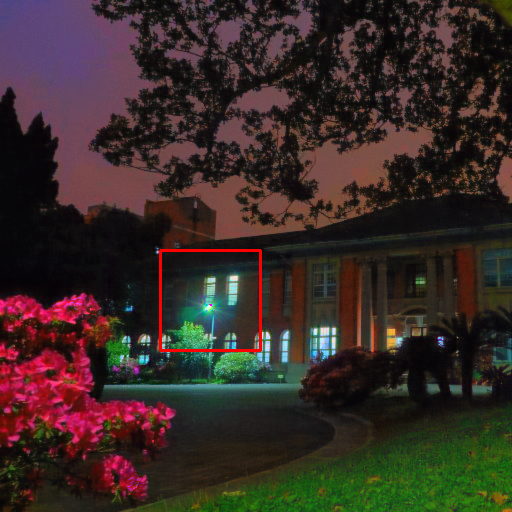}	\\
		\includegraphics[height=0.875\linewidth]{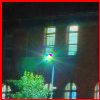}	
		\caption{SingleHDR \cite{Liu2020single}}
	\end{subfigure}
	\begin{subfigure}[b]{0.16\textwidth}
		\centering
		\includegraphics[height=0.875\linewidth]{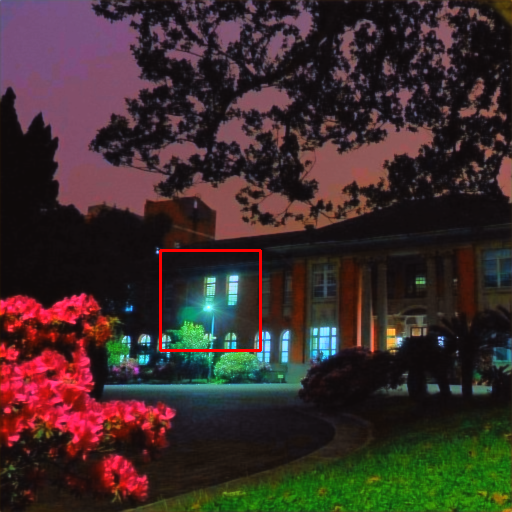}	\\
		\includegraphics[height=0.875\linewidth]{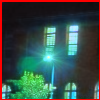}	
		\caption{Ours}
	\end{subfigure}
	\begin{subfigure}[b]{0.16\textwidth}
		\centering
		\includegraphics[height=0.875\linewidth]{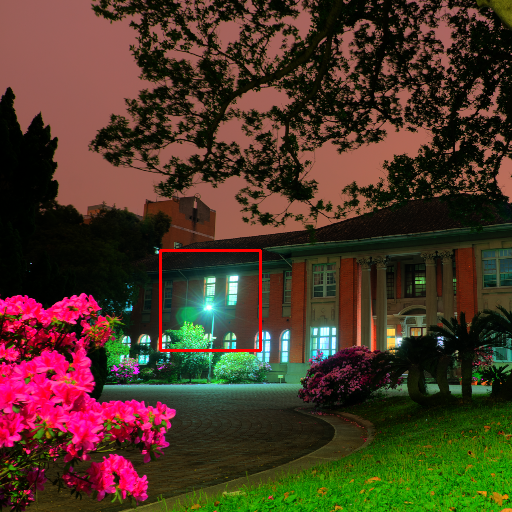}	\\
		\includegraphics[height=0.875\linewidth]{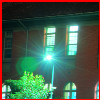}	
		\caption{Ground Truth}
	\end{subfigure}
	\caption{Visual comparison of synthesized HDR images with different algorithms. A sub-region highlighted in the red box is amplified and shown underneath each image for easy visualization. All the HDR images are tone-mapped using algorithm \cite{Drago2003}.}
	\label{fig:fig4}
	\centering
	\begin{subfigure}[b]{0.172\textwidth}
		\centering
		\includegraphics[height=0.85\linewidth]{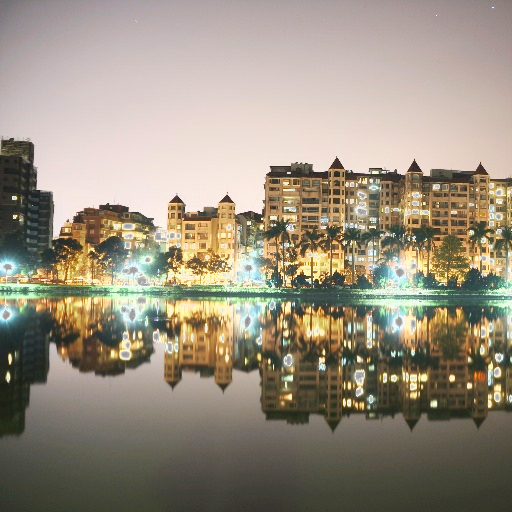}	\\
		\includegraphics[height=0.85\linewidth]{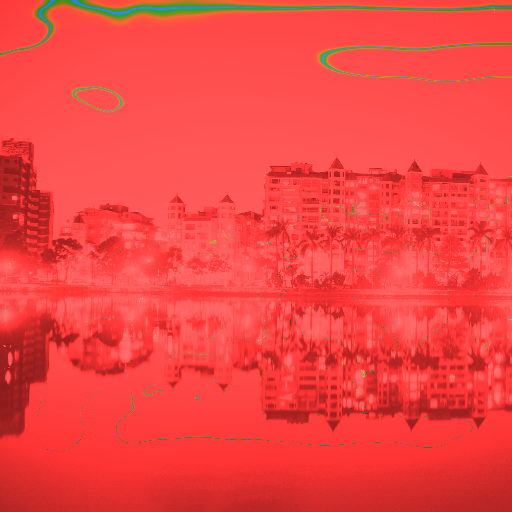}	
		\caption{DrTMO \cite{Endo2017deep}}
	\end{subfigure}	
	\begin{subfigure}[b]{0.172\textwidth}
		\centering
		\includegraphics[height=0.85\linewidth]{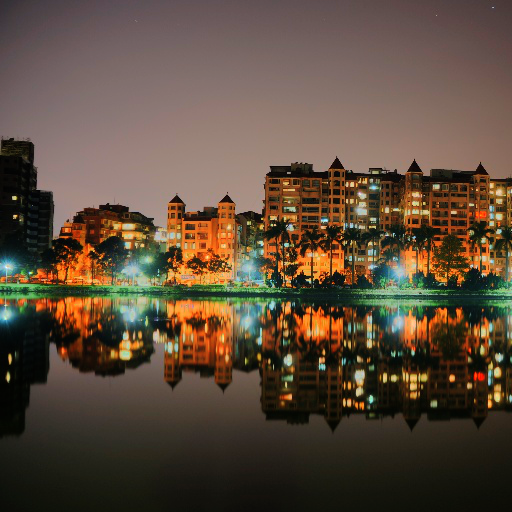} \\
		\includegraphics[height=0.85\linewidth]{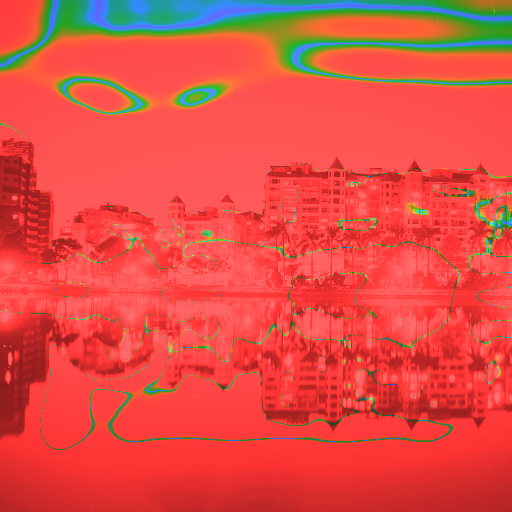}	
		\caption{HDRCNN \cite{Eilertsen2017hdr}}
	\end{subfigure}	
	\begin{subfigure}[b]{0.172\textwidth}
		\centering
		\includegraphics[height=0.85\linewidth]{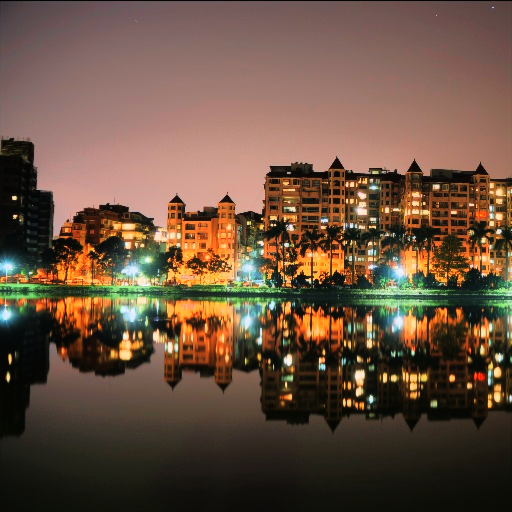}  \\
		\includegraphics[height=0.85\linewidth]{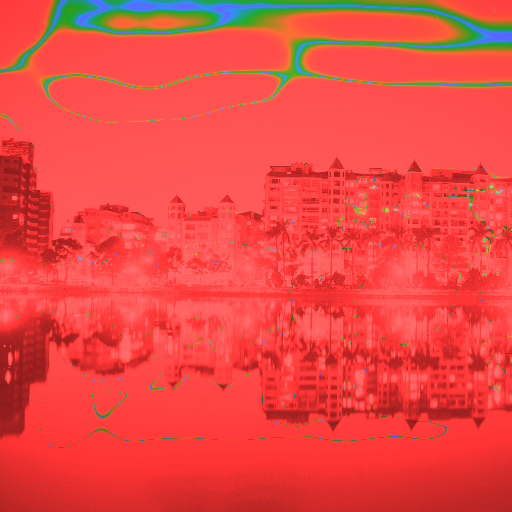}	
		\caption{ExpandNet \cite{Marnerides2018expandnet}}
	\end{subfigure}	
	\begin{subfigure}[b]{0.172\textwidth}
		\centering
		\includegraphics[height=0.85\linewidth]{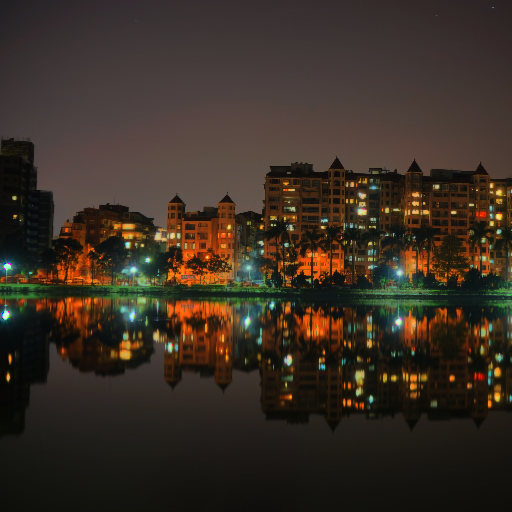}	\\
		\includegraphics[height=0.85\linewidth]{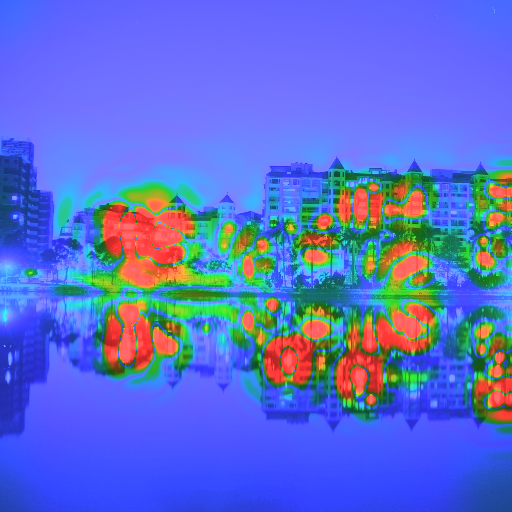}
		\caption{SingleHDR \cite{Liu2020single}}
	\end{subfigure}	
	\begin{subfigure}[b]{0.172\textwidth}
		\centering
		\includegraphics[height=0.85\linewidth]{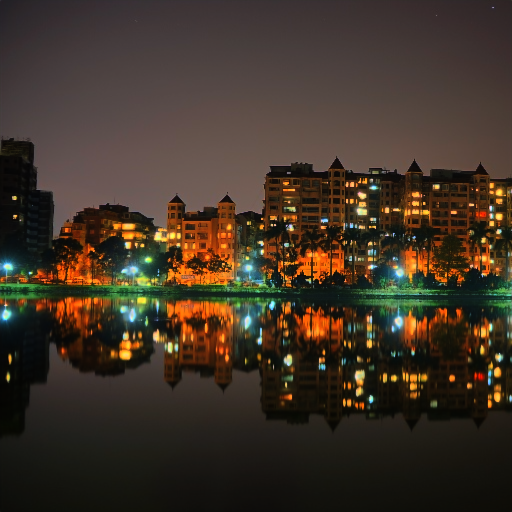}	\\
		\includegraphics[height=0.85\linewidth]{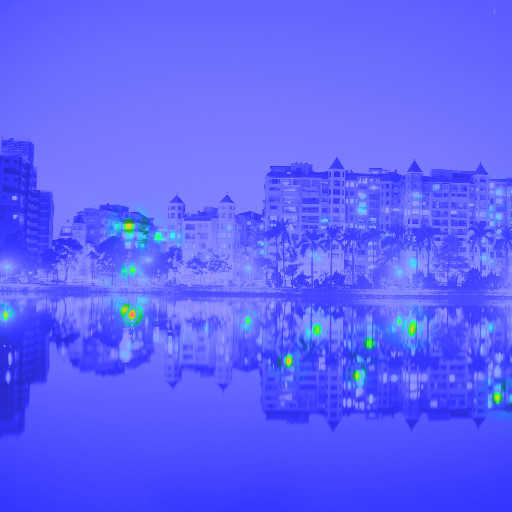}
		\caption{Ours}
	\end{subfigure}		
	\caption{Inverse tome mapping for LDR image captured at night. The upper row shows the synthesized HDR images using learning-based methods, and the bottom row shows their corresponding HDR-VDP-2 visibility probability maps.  Blue  and red indicate imperceptible and perceptible differences, respectively. All the HDR images are tone-mapped using algorithm \cite{Drago2003}.}
	\label{fig:fig5}
\end{figure*}

\subsection{Comparisons with state-of-the-art methods}
Our proposed method is compared against model-based algorithms, \textit{e.g.}, AEO \cite{Akyuz2007hdr}, HPEO  \cite{Huo2014physiological}, KOEO \cite{Kovaleski2014high} and MEO \cite{Masia2017dynamic}, as well as learning-based methods including HDRCNN \cite{Eilertsen2017hdr}, DrTMO \cite{Endo2017deep}, ExpandNet \cite{Marnerides2018expandnet} and SingleHDR \cite{Liu2020single}. 
Table \ref{t1} lists the quantitative results measured by HDR-VDP, PU-PSNR, PU-SSIM and PU-MS-SSIM. 
It is clearly seen that model-based methods have a worse performance than learning-based methods, among which the proposed method has a favorable performance against the state-of-the-art method.
Noticeably, our method has improved the HDR-VDP by 0.91, and PU-PSNR by 3.95 dB over SingleHDR on the RAISE dataset, therefore demonstrating the effectiveness of proposed method. 

Furthermore, we have conducted comprehensive visual comparisons to show the superiority of the proposed method over the existing ones.
Firstly, Fig. \ref{fig:fig4} shows the synthesized HDR images using different algorithms. 
It is clearly seen that the learning-based methods have an improved visual performance over the model-based ones, \textit{e.g.}, AEO, HPEO, KOEO and MEO, which have failed to restore the missing contents in the region marked by the red box, \textit{i.e.}, the synthesized HDR images suffer from visual artifacts.
Notably, our synthesized image shows good consistency with the ground truth one, demonstrating the favorable performance of the proposed method.

\begin{figure*}[htp]
	\centering
	\begin{subfigure}[b]{0.172\textwidth}
		\centering
		\includegraphics[height=0.85\linewidth]{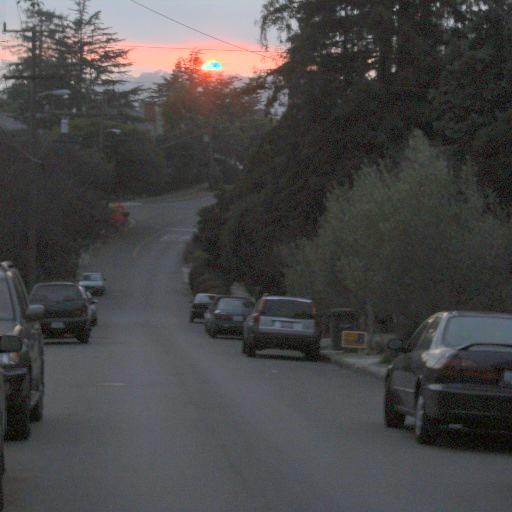}	\\
		\includegraphics[height=0.85\linewidth]{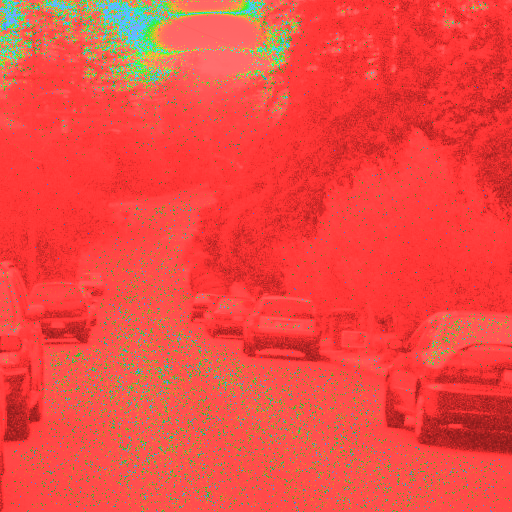}	
		\caption{DrTMO \cite{Endo2017deep}}
	\end{subfigure}	
	\begin{subfigure}[b]{0.172\textwidth}
		\centering
		\includegraphics[height=0.85\linewidth]{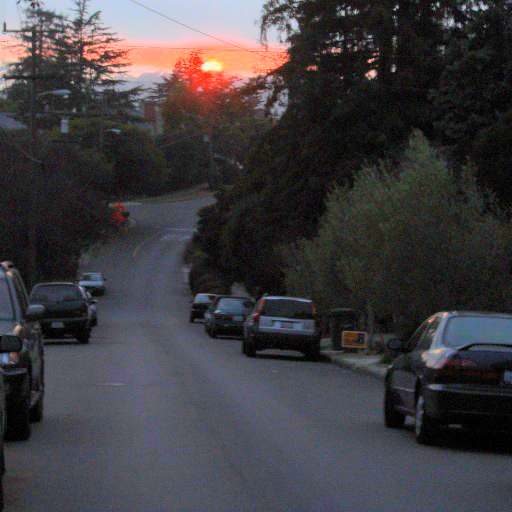} \\
		\includegraphics[height=0.85\linewidth]{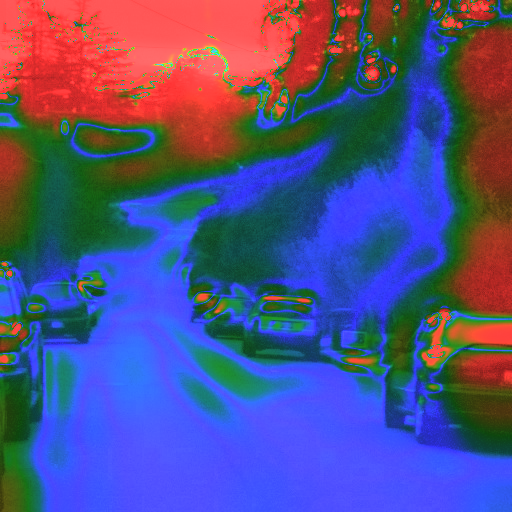}	
		\caption{HDRCNN \cite{Eilertsen2017hdr}}
	\end{subfigure}	
	\begin{subfigure}[b]{0.172\textwidth}
		\centering
		\includegraphics[height=0.85\linewidth]{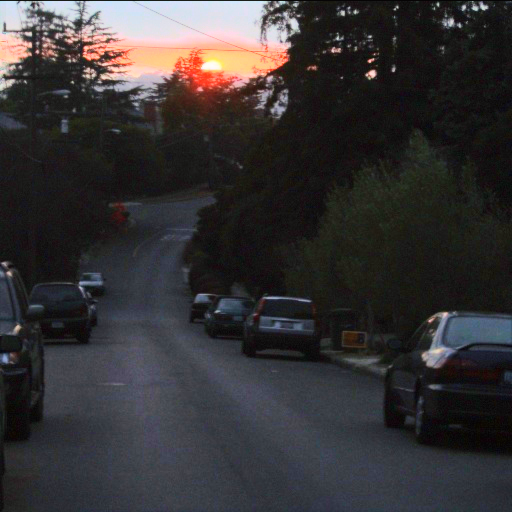}  \\
		\includegraphics[height=0.85\linewidth]{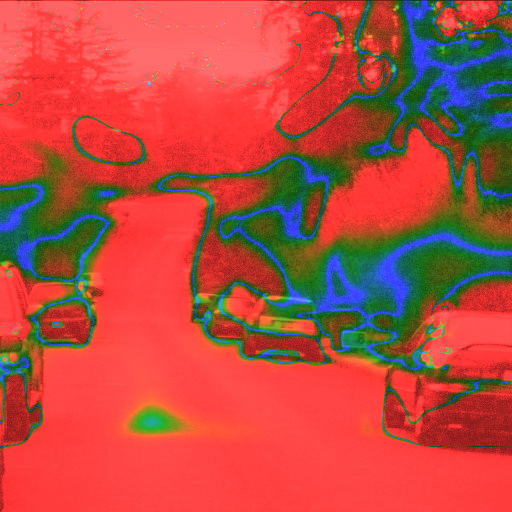}	
		\caption{ExpandNet \cite{Marnerides2018expandnet}}
	\end{subfigure}	
	\begin{subfigure}[b]{0.172\textwidth}
		\centering
		\includegraphics[height=0.85\linewidth]{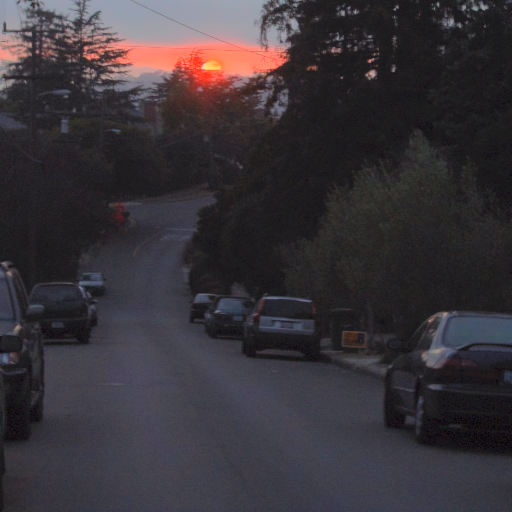}	\\
		\includegraphics[height=0.85\linewidth]{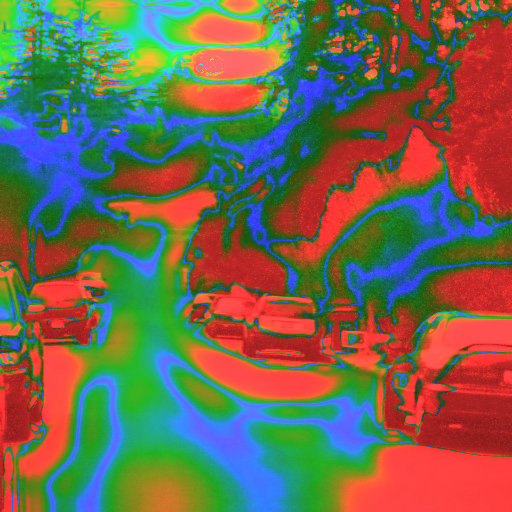}
		\caption{SingleHDR \cite{Liu2020single}}
	\end{subfigure}	
	\begin{subfigure}[b]{0.172\textwidth}
		\centering
		\includegraphics[height=0.85\linewidth]{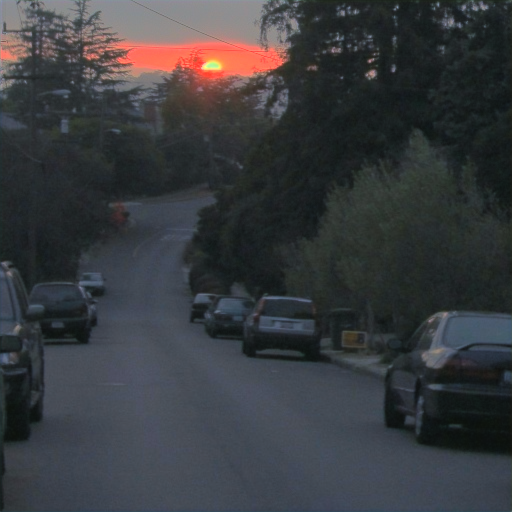}	\\
		\includegraphics[height=0.85\linewidth]{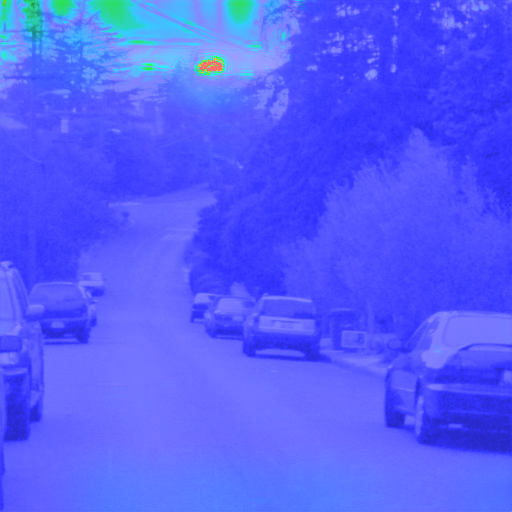}
		\caption{Ours}
	\end{subfigure}	
	\caption{Another case of inverse tome mapping with LDR input captured in the daytime.}
	\label{fig:fig6}
		\centering
	\includegraphics[height=0.5\linewidth]{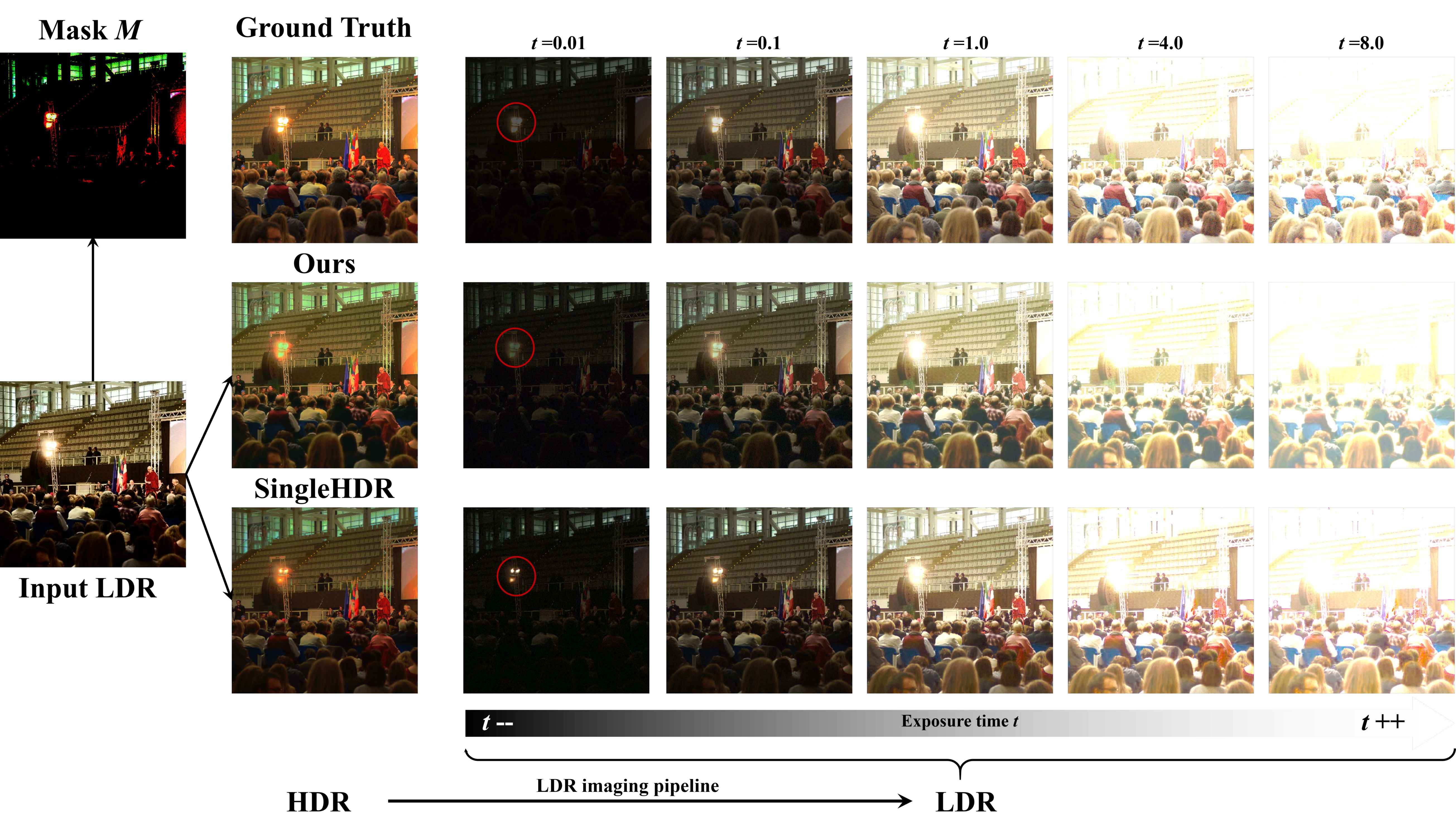}	
	\caption{Visual comparison between SingleHDR \cite{Liu2020single} and the proposed method. The first left column plots the input LDR image and its corresponding mask $M$. To clearly show the entire dynamic range in a Standard Dynamic Range (SDR) display, the LDR imaging pipeline is applied to obtain the LDR images with multiple exposures. The exposures are in the set of [0.01, 0.1, 1.0, 4.0, 8.0], and the CRF function is randomly selected from the CRF dataset. The upper row shows the ground truth HDR image and its corresponding LDR ones obtained by the LDR imaging pipeline. In the same way, the middle and bottom row are obtained using the proposed method and SingleHDR, respectively. All the HDR images are tone-mapped using algorithm \cite{Drago2003} and showed in the second left column.}
	\label{fig:fig7}
\end{figure*}

Secondly, Fig. \ref{fig:fig5} and \ref{fig:fig6} show two cases of inverse tone mapping using the learning-based methods, one with LDR input captured at night, and the other with LDR input captured in the daytime.
The visibility probability maps of HDR-VDP-2 in Fig. \ref{fig:fig5} and \ref{fig:fig6} indicate the visual quality of synthesized images.
It is rather obvious that SingleHDR has improved the visual quality over DrTMO, HDRCNN and ExpandNet, but still performs worse than the proposed method.

Traditional Standard Dynamic Range (SDR) monitor can only display brightness up to 100 nits. To clearly show the entire dynamic range in a SDR monitor, we utilize the LDR imaging pipeline to convert a synthesized HDR image to multiple LDR ones with different exposures. Thus, we can study the visual differences in the entire dynamic ranges. Using this method, a detailed comparison between the proposed method and SingleHDR is conducted and shown in Fig. \ref{fig:fig7}.
As indicated by the mask $M$ in Fig. \ref{fig:fig7},  sensor saturation has caused dynamic range clipping in the 'lamp' regions. It is clearly seen that SingleHDR \cite{Liu2020single} has failed to infer the missing contents in the lamp region as there exists clear differences between the synthesized images of SingleHDR and that of the ground truth (highlighted by red circles in Fig. \ref{fig:fig7}). In contrast, our synthesized HDR image is consistent with the ground truth, especially in the over-exposed regions.

\begin{table*}[htp]
	\centering
	\caption{Performance evaluation with different settings. Config A hierarchically synthesizes $h_1^{'}$, $h_2^{'}$ and $h_3^{'}$, which corresponds to a particular inverse process of the LDR imaging pipeline, \textit{i.e.}, the quantization, CRF and dynamic range clipping process, respectively. Config B removes the hierarchical synthesis process, and generates the HDR image directly. The bold indicates the best result.}
	\resizebox{170mm}{10mm}{
		\begin{tabular}{l|c c c c||c c c c|| c c c c}
			\hline\hline
			&\multicolumn{4}{c||}{HDR-SYNTH}&\multicolumn{4}{c||}{HDR-REAL}&\multicolumn{4}{c}{RAISE}	\\
			& {\scriptsize HDR-VDP} & {\scriptsize PU-PSNR} & {\scriptsize PU-SSIM} & {\scriptsize PU-MSSSIM} & {\scriptsize HDR-VDP} & {\scriptsize PU-PSNR} & {\scriptsize PU-SSIM} & {\scriptsize PU-MS-SSIM} & {\scriptsize HDR-VDP} & {\scriptsize PU-PSNR} & {\scriptsize PU-SSIM} & {\scriptsize PU-MS-SSIM}\\
			\hline
			Config A&56.87 & 28.55 & 0.51 & 0.44 & 60.10 & 36.15 &0.69 & 0.64 & 59.99 & 33.11 &0.62 & 0.56\\
			Config B& 51.67 & 23.41 &0.37 &0.29 & 58.61 & 31.77 & 0.57 & 0.56 & 58.11 &32.84 & 0.51 & 0.47\\		
			Ours&\textbf{57.46}&\textbf{29.3}0&\textbf{0.67}&\textbf{0.58}&\textbf{61.33}&\textbf{37.95}&\textbf{0.73}&\textbf{0.70}&\textbf{60.79}&\textbf{34.43}&\textbf{0.71}&\textbf{0.68}\\
			\hline\hline
	\end{tabular}}
	\label{t2}
\end{table*}
\begin{figure*}[htp]
	\centering
	\begin{subfigure}[b]{0.1\textwidth}
		\centering
		\includegraphics[height=1.0\linewidth]{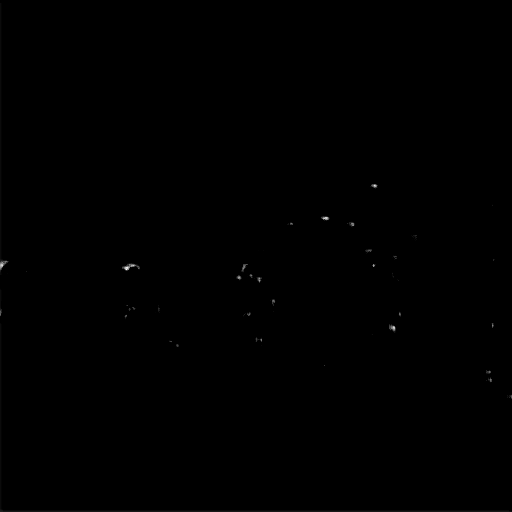}	
		\caption{}
	\end{subfigure}	
	\begin{subfigure}[b]{0.1\textwidth}
		\centering
		\includegraphics[height=1.0\linewidth]{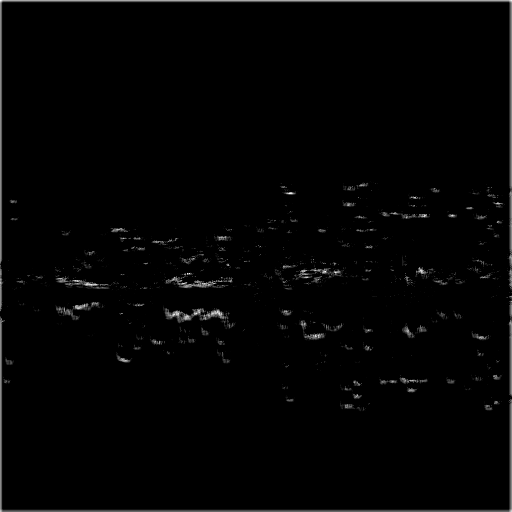}	
		\caption{}
	\end{subfigure}
	\begin{subfigure}[b]{0.1\textwidth}
		\centering
		\includegraphics[height=1.0\linewidth]{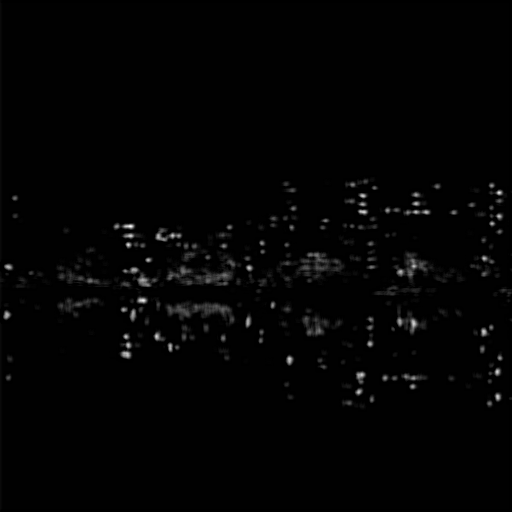}	
		\caption{}
	\end{subfigure}
	\begin{subfigure}[b]{0.1\textwidth}
		\centering
		\includegraphics[height=1.0\linewidth]{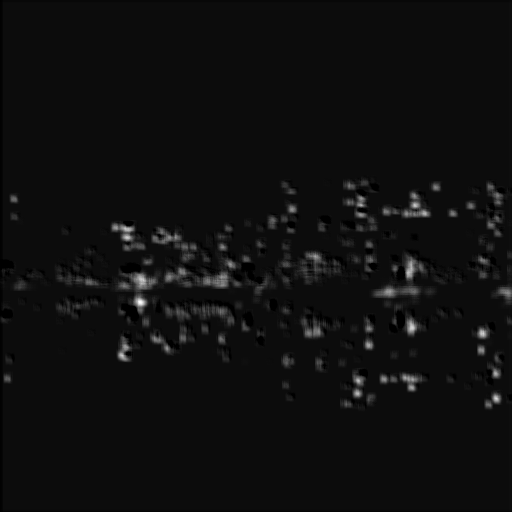}	
		\caption{}
	\end{subfigure}
	\begin{subfigure}[b]{0.1\textwidth}
		\centering
		\includegraphics[height=1.0\linewidth]{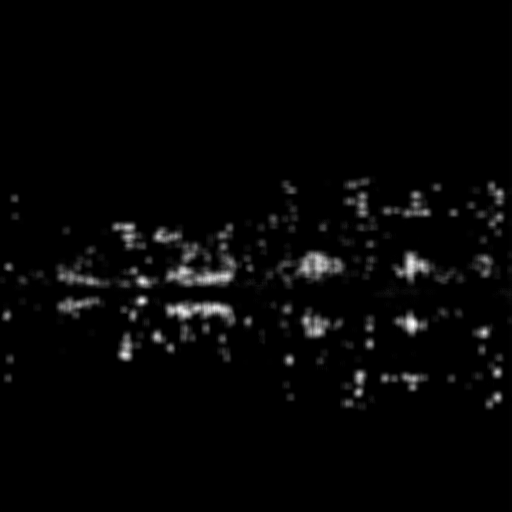}	
		\caption{}
	\end{subfigure}
	\begin{subfigure}[b]{0.1\textwidth}
		\centering
		\includegraphics[height=1.0\linewidth]{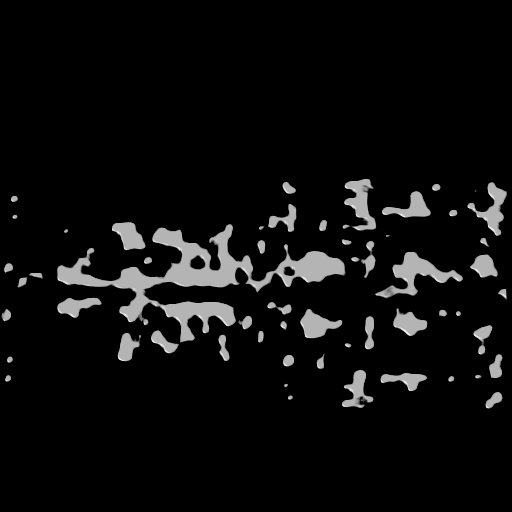}	
		\caption{}
	\end{subfigure}
	\begin{subfigure}[b]{0.1\textwidth}
		\centering
		\includegraphics[height=1.0\linewidth]{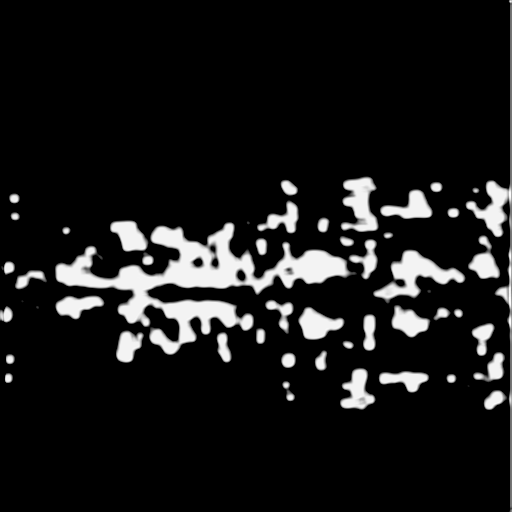}	
		\caption{}
	\end{subfigure}
	\begin{subfigure}[b]{0.1\textwidth}
		\centering
		\includegraphics[height=1.0\linewidth]{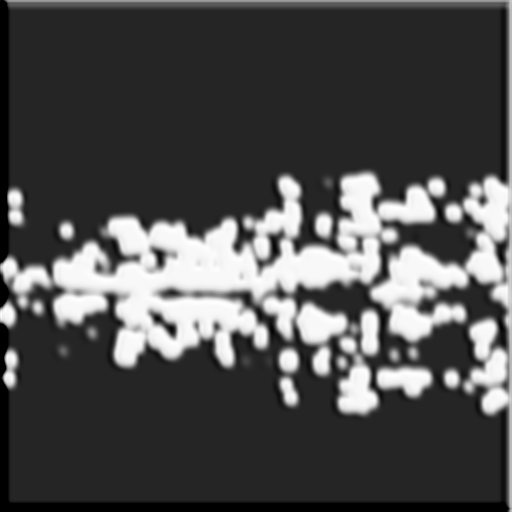}	
		\caption{}
	\end{subfigure}
	\begin{subfigure}[b]{0.1\textwidth}
		\centering
		\includegraphics[height=1.0\linewidth]{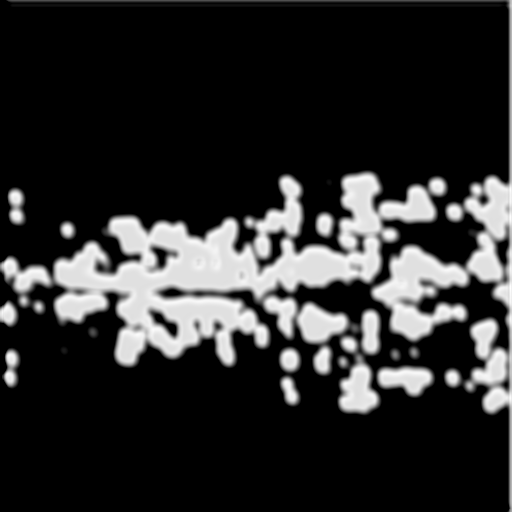}	
		\caption{}
	\end{subfigure}
	
	\begin{subfigure}[b]{0.1\textwidth}
		\centering
		\includegraphics[height=1.0\linewidth]{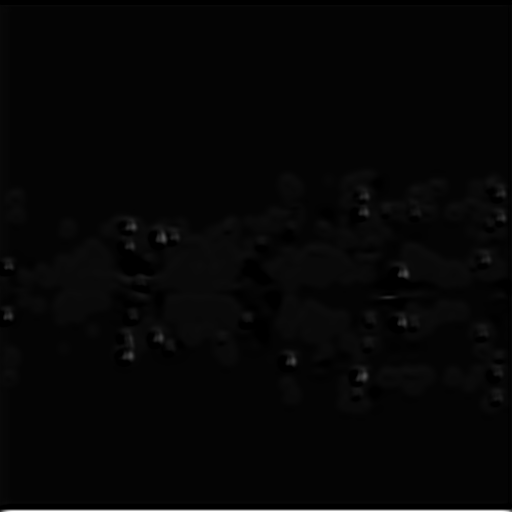}	
		\caption{}
	\end{subfigure}
	\begin{subfigure}[b]{0.1\textwidth}
		\centering
		\includegraphics[height=1.0\linewidth]{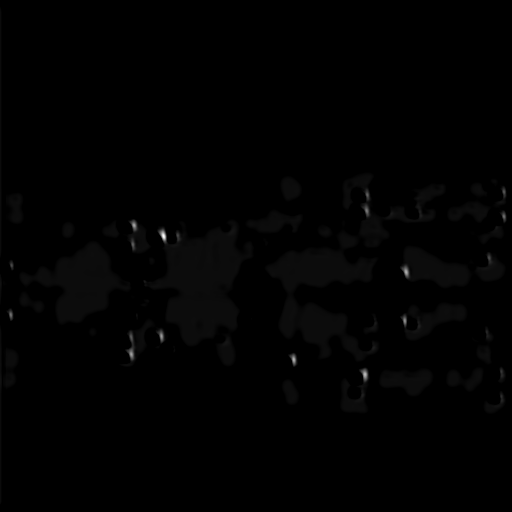}	
		\caption{}
	\end{subfigure}
	\begin{subfigure}[b]{0.1\textwidth}
		\centering
		\includegraphics[height=1.0\linewidth]{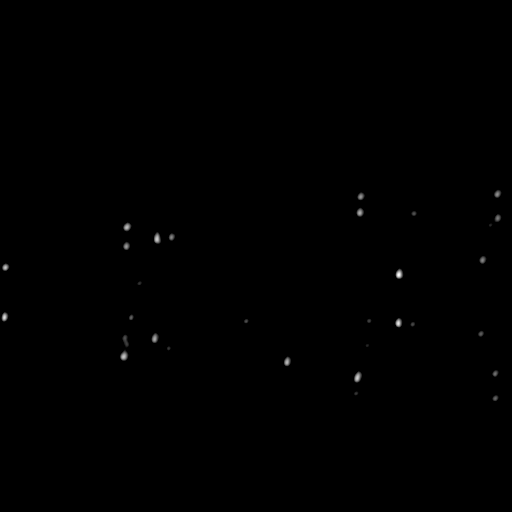}	
		\caption{}
	\end{subfigure}
	\begin{subfigure}[b]{0.1\textwidth}
		\centering
		\includegraphics[height=1.0\linewidth]{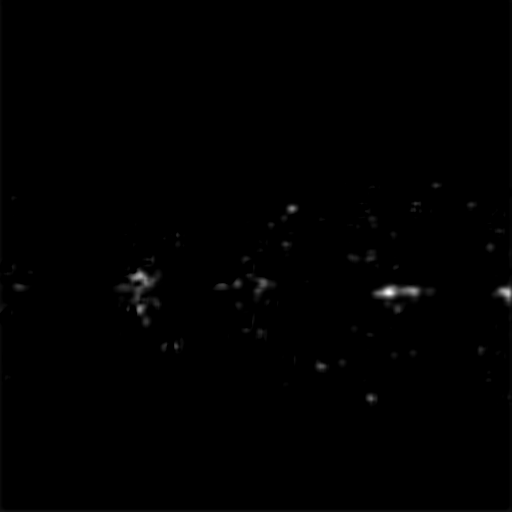}	
		\caption{}
	\end{subfigure}
	\begin{subfigure}[b]{0.1\textwidth}
		\centering
		\includegraphics[height=1.0\linewidth]{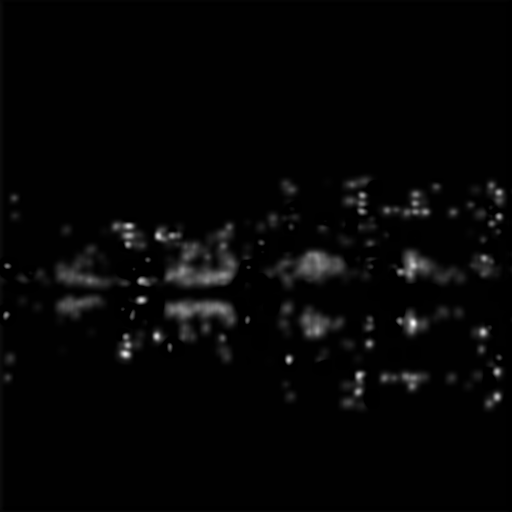}	
		\caption{}
	\end{subfigure}
	\begin{subfigure}[b]{0.1\textwidth}
		\centering
		\includegraphics[height=1.0\linewidth]{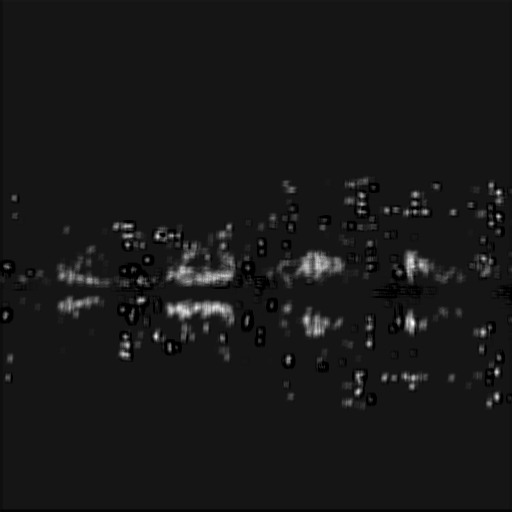}	
		\caption{}
	\end{subfigure}
	\begin{subfigure}[b]{0.1\textwidth}
		\centering
		\includegraphics[height=1.0\linewidth]{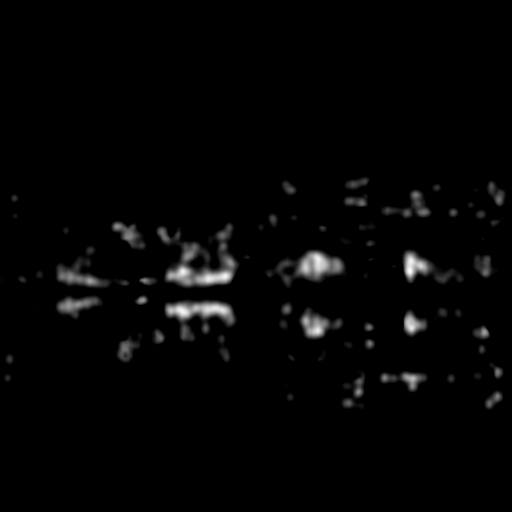}	
		\caption{}
	\end{subfigure}
	\begin{subfigure}[b]{0.1\textwidth}
		\centering
		\includegraphics[height=1.0\linewidth]{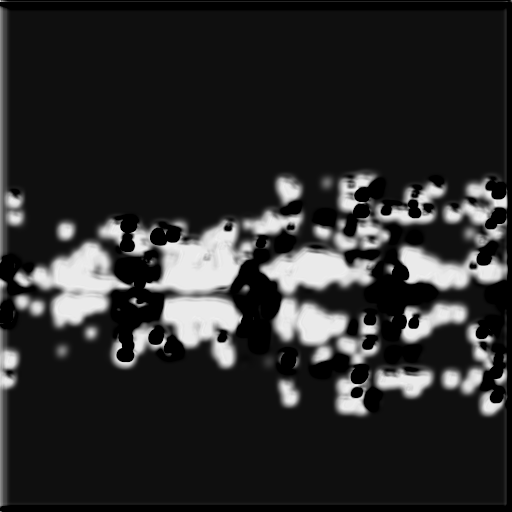}	
		\caption{}
	\end{subfigure}
	\begin{subfigure}[b]{0.1\textwidth}
		\centering
		\includegraphics[height=1.0\linewidth]{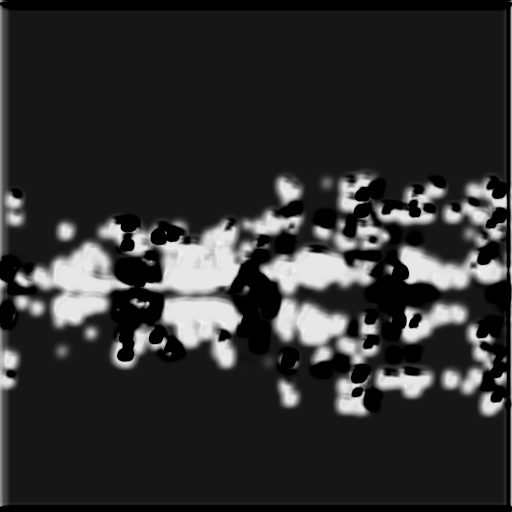}	
		\caption{}
	\end{subfigure}	
	\caption{Illustration of  the activation maps of $\gamma_i$ and $\beta_i$. (a)-(i)  show part of the activation maps in $\gamma_i$, \textit{e.g.}, (a) plots the $1$-th channel of $\gamma_6$. In the same way, (j)-(r) show activation maps of $\beta_i$.}
	\label{fig:fig8}
\end{figure*}

The quantitative results listed in Table \ref{t1} as well as the visual comparisons present in Fig. \ref{fig:fig4}, \ref{fig:fig5}, \ref{fig:fig6} and \ref{fig:fig7} have demonstrated the improved performance of the proposed method over the state-of-the-art algorithms on the iTM tasks. To explain how the proposed method works, ablation studies are conducted and described as followed.

\section{Ablation Studies}
\subsection{Design of HiSN}\label{hisn}
HiSN incorporates the prior knowledge of the LDR imaging pipeline in the HDR synthesis process.
To show the effectiveness of HiSN, comparative trials are conducted. To be specific, Config A hierarchically synthesizes $h_1^{'}$,  $h_2^{'}$ and  $h_3^{'}$. Each one of the three parts is responsible for one particular inverse process of the LDR imaging pipeline, with $h_1^{'}$ corresponding to the inverse quantization process, $h_2^{'}$ for the inverse CRF process and $h_3^{'}$ for the inverse dynamic range clipping process.
Thus, Config A synthesizes the HDR outputs via: $L \rightarrow h_1^{'} \rightarrow h_2^{'} \rightarrow h_3^{'}$, in contrast to the proposed HiSN design:  $L \rightarrow H_1 \rightarrow H_2$.
As a comparison, Config B removes the hierarchical synthesis process and generate the HDR output directly: $L \rightarrow H$.
Performance evaluation with different settings is listed in Table \ref{t2}.
Firstly, both Config A and Config B  have observed decreased performances measured by HDR-VDP, PU-PSNR, PU-SSIM and PU-MS-SSIM.
Secondly, Config B performs worse than Config A and the proposed method, indicating that the hierarchical synthesis process indeed contributes to alleviating the difficulty of the LDR-to-HDR mapping function.
Thirdly, Config A conducts the hierarchical synthesis process in a reverse order of the LDR imaging pipeline (Fig. \ref{fig:fig2}),
However, Config A compromises the iTM performance instead of improving it. Empirically, utilizing $H_1$ to conduct radiometric calibration and infer the missing contents caused by quantization leads to a better solution, demonstrating the effectiveness of our strategy of separating the image regions based on their lightness.

\subsection{How does LAMN work? }\label{mamn}
Both visual and quantitative results have demonstrated the favorable performance of the proposed method, but the question why and how LAMN contributes to inferring the missing contents in the over-exposed regions still persists.

\begin{figure*}[htp]
	\centering
	\includegraphics[height=0.4\linewidth]{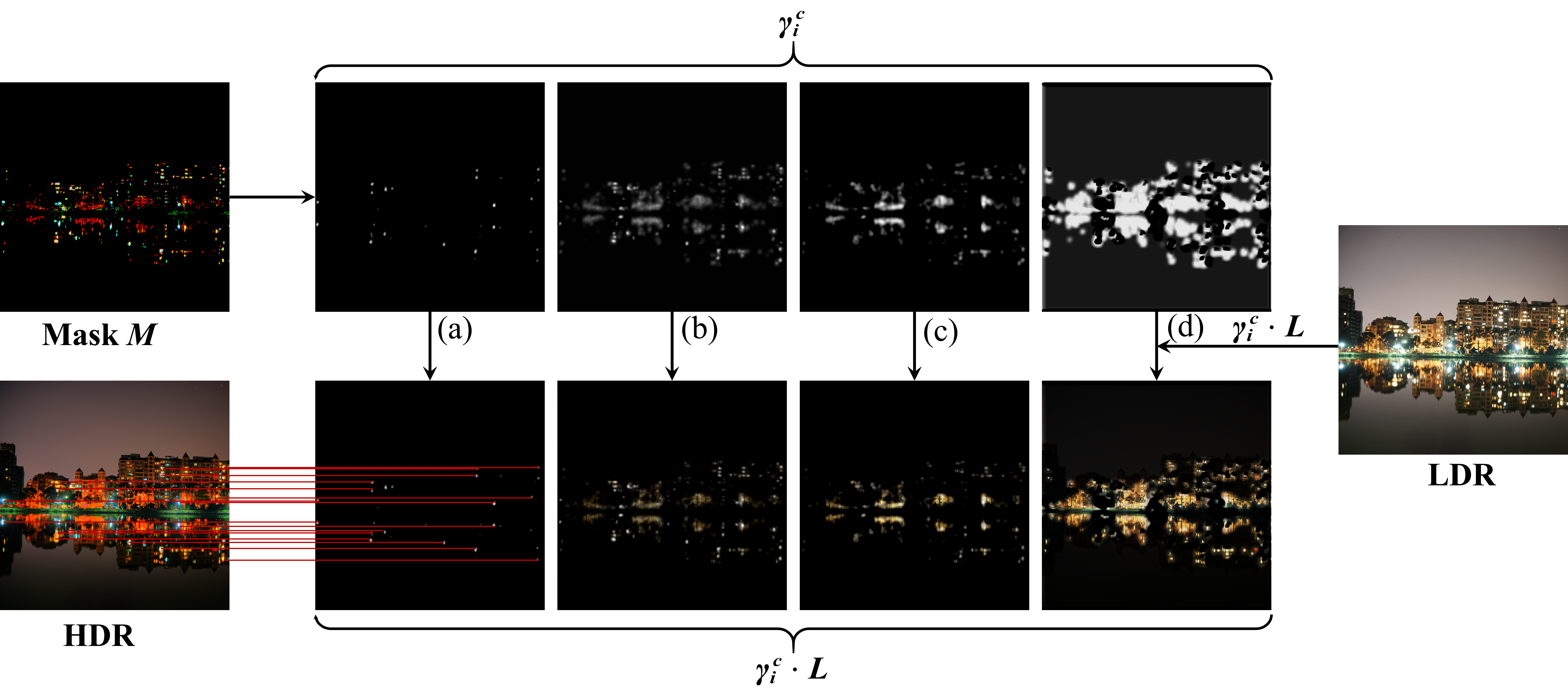}	
	\caption{Illustration of the effect of $\gamma_i$ on the feature maps. The red lines plot the correspondences between the highlighted parts of $\gamma_i^{n}\cdot L$ and the HDR reference. It is clear that the highlighted parts in  $\gamma_i^{n}\cdot L$ correspond to regions with highest irradiance intensities in the HDR reference.}
	\label{fig:fig9}
	\centering
	\begin{subfigure}[b]{0.24\textwidth}
		\centering
		\includegraphics[height=0.9\linewidth]{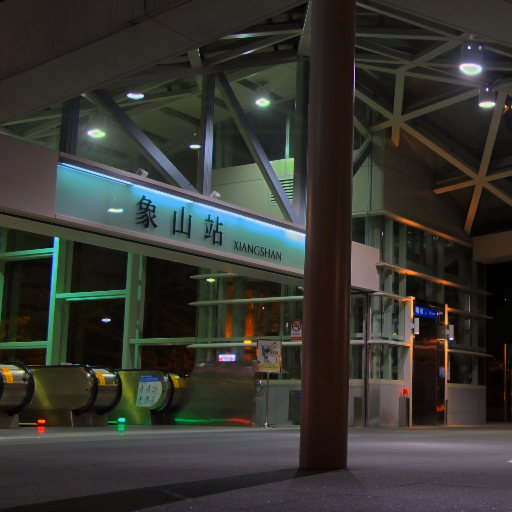}	\\
		\includegraphics[height=0.9\linewidth]{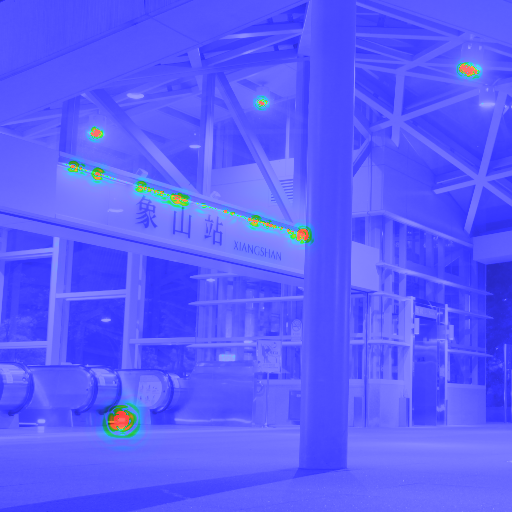}	
		\caption{Ours}
	\end{subfigure}	
	\begin{subfigure}[b]{0.24\textwidth}
		\centering
		\includegraphics[height=0.9\linewidth]{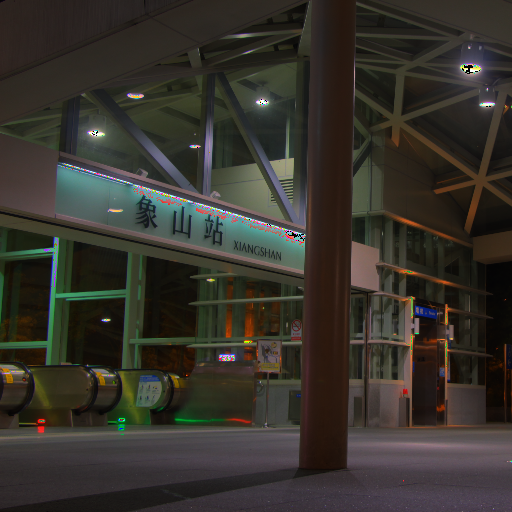}	\\
		\includegraphics[height=0.9\linewidth]{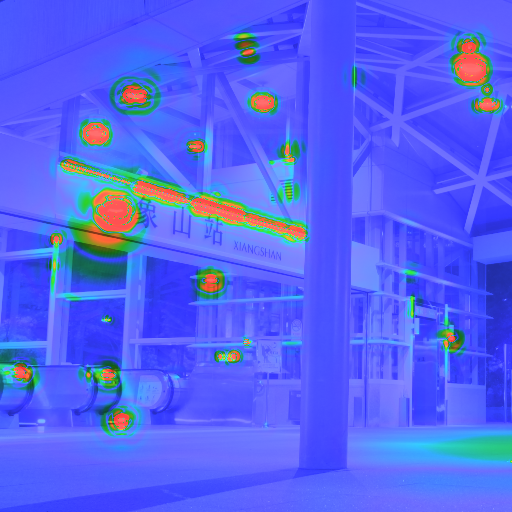}	
		\caption{Config C}
	\end{subfigure}	
	\begin{subfigure}[b]{0.24\textwidth}
		\centering
		\includegraphics[height=0.9\linewidth]{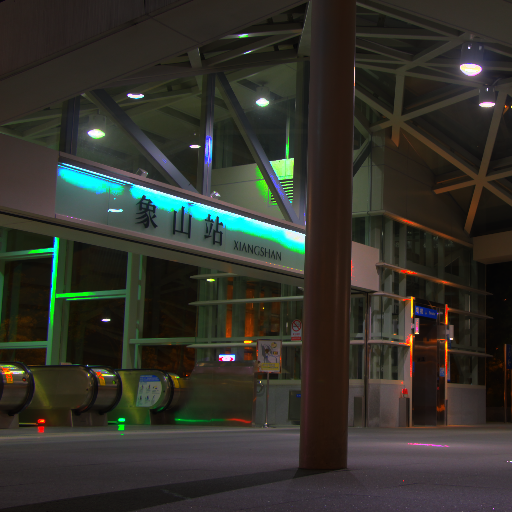}	\\
		\includegraphics[height=0.9\linewidth]{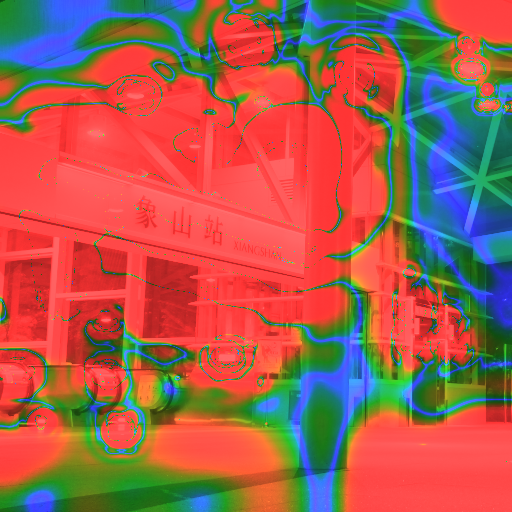}	
		\caption{Config D}
	\end{subfigure}	
	\begin{subfigure}[b]{0.24\textwidth}
		\centering
		\includegraphics[height=0.9\linewidth]{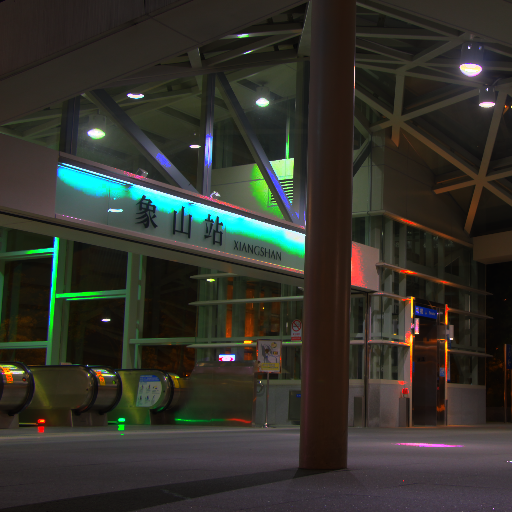}	\\
		\includegraphics[height=0.9\linewidth]{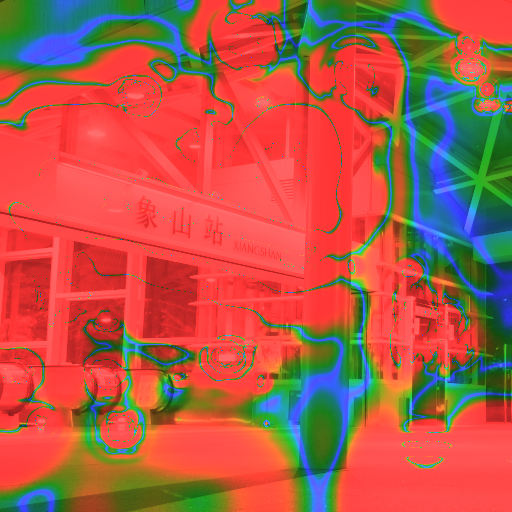}	
		\caption{Config E}
	\end{subfigure}	
	\caption{An example showing the effectiveness of the mask $M$. The upper row shows synthesized HDR images with different settings, and the bottom indicates their corresponding HDR-VDP visibility probability maps. (a)-(d) show the results using our proposed method, Config C, Config D and Config E, respectively.}
	\label{fig:fig10}
\end{figure*}

To gain an insight into the working of LAMN, we set out to investigate the channels in  $\gamma_i \in R^{C_i \times H_i \times W_i}$ and $\beta_i \in R^{C_i \times H_i \times W_i}$, \textit{e.g.}, $\gamma_i^{n} \in R^{H_i \times W_i}$, which represents the $n$-th channel in $\gamma_i$ ($n \in [0, C_i]$). 
$\gamma_i^{n}$ and $\beta_i^{n}$ are normalized and then shown as gray-scale images in Fig. \ref{fig:fig8}.
It is clearly seen that $\gamma_i^{n}$ and $\beta_i^{n}$ would affect different areas of the feature maps. 
Furthermore, Fig. \ref{fig:fig9} shows the effect of $\gamma_i$ on the feature maps. Specifically, we utilize the LDR image as a representation of the feature map, and channel-wisely compute $\gamma_i^{n}\cdot L$  to investigate which regions will be affected by $\gamma_i$.
As indicated by Fig. \ref{fig:fig8} and \ref{fig:fig9}, $\gamma_i$ and $\beta_i$ would modulate the over-exposed regions, and hardly affected the under-exposed area. 
In addition, $\gamma_i^{n}$ of  (a) in Fig. \ref{fig:fig9} would enhance features whose positions correspond to the highest irradiance intensities in the HDR reference. As a comparison, $\gamma_i^{n}$ of  (d) would affect most of the over-exposed regions.
The dynamic range clipping process would result in all the pixel values in the over-exposed regions being clipped to 1, and there exists no variations.
Through the modulation of LAMN, the features, whose positions correspond to higher irradiance intensities, would more likely to be modulated, \textit{e.g.}, the case of (a) in Fig. \ref{fig:fig9}, while those corresponding to lower irradiance intensity would less likely to be modulated, like the case in Fig. \ref{fig:fig9} (d). Thus, the learned parameters $\gamma_i$ and $\beta_i$ would adaptively and discriminately modulate the features. This is achieved by bringing in pixels in the areas surrounding the over-exposed regions into the process of estimating the missing details in the saturated areas via LAMN.
Conventional convolutional operations are conducted in a sliding window manner, handling all the pixels equally. 
As a comparison, LAMN would modulate the features adaptively and discriminately via the learned parameters based on the lightness adaptive mask $M$. Thus, HiSN process different lightness regions differently rather than treating them equally.
Considering the quantitative and visual results, the modulation technique in LAMN indeed contributes to inferring the missing contents in the over-exposed region.

Moreover, we have observed that the mask $M$, acting as prior knowledge to indicate $P_s$ and $P_a$, contributes to the iTM performance.
As indicated by Fig. \ref{fig:fig3}, $P_a$ shows greater relevance to inferring the saturated pixel $P_s$, and has been taken into consideration in $M$, thus intuitively helping HiSN focus on features having close relationship to $P_s$ and $P_a$, and further contributing to the iTM performance. 
To elaborate on this, comparative trials are conducted and results are shown in Fig. \ref{fig:fig10}. Specifically, Config C takes $\tau$ in (\ref{key3}) as $1-1e^{-10}$ in the HDR synthesis process, \textit{i.e.}, no $P_a$ is considered in the mask $M$, and each entry of the mask $M$ in Config D and Config E is set as 0 and 1, respectively, \textit{i.e.}, the mask $M$ has no effect on $x_i$ in Config D, and Config E utilizes all the pixels equally to help HiSN infer the saturated regions.
It is clearly seen that removing $P_a$ in the mask $M$ of Config C leads to the decreased performance in Fig. \ref{fig:fig10} (b), and Config C suffers from perceptible differences in the over-exposed regions. Meanwhile, Config D and Config E show two cases where the mask $M$ are used improperly.
$M$ would fail to indicate the relationship between $P_s$ and $P_a$ when setting all the entries in $M$ to 0 or 1. Thus, the synthesized HDR images would have poor visual qualities in Fig. \ref{fig:fig10} (c) and (d)

To summarize, interpretation of LAMN module can be given as followed. The mask $M$ can be regarded as prior knowledge to indicate $P_s$, $P_a$ and their relationships. Then, (\ref{key4}) provides a technique for modulating the feature maps in HiSN, such that $(1+\gamma_i)$ and $\beta_i$ could adaptively and discriminatively scale and bias the feature $x_i$, respectively.
Eventually, HiSN is able to focus on relevant features when inferring the scene irradiance of the over-exposed regions. 


%
%
%

\section{Conclusions}
To conduct inverse tone mapping, we make use of the lightness prior and have introduced  an novel architecture consisting of the Hierarchical Synthesis Network (HiSN) and the Lightness Adaptive Modulation Network (LAMN). To obtain the HDR estimation, HiSN synthesize the dim part first, and generates the bright part with the modulation of LAMN. LAMN can adaptively and discriminately modulate the features in HiSN, enabling HiSN to focus on relevant feature, thus inferring the missing contents in the over-exposed region can be achieved in a better way. Extensive experiments have confirmed the effectiveness of HiSN and LAMN.
Furthermore, we visualize the activation maps of $\gamma_i$ and $\beta_i$ to investigate how modulation is conducted, and analyze how inferring the saturated pixels is achieved via LAMN. 
Overall, the proposed method has observed improved performance over state-of-the art algorithms on the inverse tone mapping task.


%

%
%
%
%
%
%

\ifCLASSOPTIONcaptionsoff
  \newpage
\fi

\bibliographystyle{IEEEtran}
\bibliography{IEEEexample}
\end{document}